\documentclass[a4paper,UKenglish]{lipics} %

\usepackage{microtype}%

\serieslogo{}%
\volumeinfo%
  {}%
  {1}%
  {}%
  {1}%
  {1}%
  {1}%
\EventShortName{}
\DOI{10.4230/LIPIcs.xxx.yyy.p}%
\title{Dynamic Graph Queries\footnote{Parts of this work are also included in the dissertation thesis of the third author \cite{Zeume15thesis}.}}

\author[1]{Pablo Mu\~{n}oz\footnote{The author acknowledges the financial support by Conicyt PhD scholarship and Millennium Nucleus Center for Semantic Web Research under Grant NC120004.}}
\author[2]{Nils Vortmeier}

\author[2]{Thomas Zeume\footnote{The author acknowledges the financial support by DFG grant SCHW 678/6-1.}}

\affil[1]{Department of Computer Science, University of Chile\\ CIWS  Center for Semantic Web Research\\
  Chile\\
  \texttt{pmunoz@dcc.uchile.cl}}

\affil[2]{TU Dortmund University\\
  Germany\\
  \texttt{\{nils.vortmeier, thomas.zeume\}@tu-dortmund.de}}

  \authorrunning{P. Mu\~{n}oz, N. Vortmeier, T. Zeume} %

\Copyright{Pablo Mu\~{n}oz, Nils Vortmeier and Thomas Zeume}%


\usepackage{amssymb,amsmath,xspace,enumerate,gensymb}
\usepackage{amsthm}

\usepackage[pagewise]{lineno}
\usepackage{graphicx}
\usepackage{xspace}
\usepackage{rotating}
\usepackage{listings}
\usepackage{soul}
\usepackage{ifmtarg}
\usepackage{stmaryrd} %
\usepackage{paralist}
\usepackage{chngcntr}
\usepackage{subfig}
\usepackage{tikz}
\usetikzlibrary{arrows,decorations.pathmorphing,backgrounds,calc,positioning,fit,petri,matrix,backgrounds, decorations.pathreplacing, shapes.geometric}
\usepackage{pgffor}
\usepackage{extarrows}
\usepackage{enumitem}
\usepackage{appendix}
\usepackage{hyperref}
\usepackage{algorithm}
\usepackage{algpseudocode}
\usepackage{array}

\newif\ifcomments
\newif\ifchanges
\commentsfalse\changesfalse
\commentstrue\changestrue %

\begin{document}
\makeatletter{}%
\makeatletter{}%
\newcommand  {\myclass} [1]  {\ensuremath{\textsc{#1}\xspace}}

\newcommand{\StaClass}[1]{\myclass{#1}\xspace}

\newcommand{\DynClass}[1]{\myclass{Dyn#1}\xspace}
\newcommand{\dDynClass}[1]{\myclass{$\Delta$-Dyn#1}\xspace}

\newcommand  {\myproblem} [1] {\textsc{#1}}

\newcommand{\problemIndent}{\hspace{5mm}}
\newcommand  {\problemdescr} [3] {
    \vspace{3mm}
    \def\Name{#1}
    \def\Input{#2}
    \def\Question{#3}
     \problemIndent\begin{tabular}{r p{\columnWidth}r}%
      \textit{Problem:} & \myproblem{\Name} \\
      \textit{Input:} & \Input \\
      \textit{Question:} & \Question
     \end{tabular}
    \vspace{3mm}
    }

\newcommand  {\querydescr} [3] {
\vspace{3mm}
\def\Name{#1}
\def\Input{#2}
\def\Question{#3}
  \problemIndent\begin{tabular}{r p{\columnWidth}r}%
  \textit{Query:} & \myproblem{\Name} \\
  \textit{Input:} & \Input \\
  \textit{Question:} & \Question
  \end{tabular}
\vspace{3mm}
}

\newcommand  {\dynproblemdescr} [4] {
    \vspace{3mm}
    \def\Name{#1}
    \def\Input{#2}
    \def\Updates{#3}  
    \def\Question{#4}
    \problemIndent\begin{tabular}{r p{\columnWidth}r}%
      \textit{Query:} & \myproblem{\Name} \\
      \textit{Input:} & \Input \\
      \textit{Question:} & \Question
    \end{tabular}
    \vspace{3mm}
}
\newcommand{\dynProbDescr}[4]{\dynproblemdescr{#1}{#2}{#3}{#4}}

\newcommand  {\problem}[1] {\myproblem{#1}}

\newcommand{\dynProb}[1] {\myproblem{Dyn(#1)}}
\newcommand{\class}{\calC}

\newcommand  {\TIME}    {\myclass{TIME}}
\newcommand  {\DLOGTIME}    {\myclass{DLogtime}}
\newcommand  {\DTIME}   {\myclass{DTIME}}
\newcommand  {\NTIME}   {\myclass{NTIME}}
\newcommand  {\ATIME}   {\myclass{ATIME}}
\newcommand  {\SPACE}   {\myclass{SPACE}}
\newcommand  {\DSPACE}   {\myclass{DSPACE}}
\newcommand  {\NSPACE}  {\myclass{NSPACE}}
\newcommand  {\coNSPACE}        {\myclass{coNSPACE}}

\newcommand     {\LOGCFL}     {\myclass{LOGCFL}}
\newcommand     {\LOGDCFL}     {\myclass{LOGDCFL}}
\newcommand     {\LOGSPACE}     {\myclass{LOGSPACE}}
\newcommand     {\NLOGSPACE}     {\myclass{NLOGSPACE}}
\newcommand     {\classL}   {\myclass{L}}
\newcommand     {\NL}   {\myclass{NL}}
\newcommand     {\coNL}   {\myclass{coNL}}
\renewcommand   {\P}    {\myclass{P}}
\newcommand     {\myP}    {\myclass{P}}
\newcommand     {\PTIME}    {\myclass{PTime}}
\newcommand     {\NP}   {\myclass{NP}}
\newcommand     {\NPC}   {\myclass{NPC}}
\newcommand     {\PH}   {\myclass{PH}}
\newcommand     {\coNP} {\myclass{coNP}}
\newcommand     {\NPSPACE}      {\myclass{NPSPACE}}
\newcommand     {\PSPACE}       {\myclass{PSPACE}}
\newcommand     {\IP}   {\myclass{IP}}
\newcommand     {\POLYLOGSPACE} {\myclass{POLYLOGSPACE}}
\newcommand     {\DET}  {\myclass{DET}}
\newcommand     {\EXP}  {\myclass{EXP}}
\newcommand     {\NEXP}  {\myclass{NEXP}}
\newcommand     {\EXPTIME}  {\myclass{EXPTIME}}
\newcommand     {\TWOEXPTIME}  {\myclass{2-EXPTIME}}
\newcommand     {\TWOEXP}  {\myclass{2-EXP}}
\newcommand     {\NEXPTIME}  {\myclass{NEXPTIME}}
\newcommand     {\coNEXPTIME}  {\myclass{coNEXPTIME}}
\newcommand     {\EXPSPACE}  {\myclass{EXPSPACE}}
\newcommand     {\RP}   {\myclass{RP}}
\newcommand     {\RL}   {\myclass{RL}}
\newcommand     {\coRP} {\myclass{coRP}}
\newcommand     {\ZPP}  {\myclass{ZPP}}
\newcommand     {\BPP}  {\myclass{BPP}}
\newcommand     {\PP}   {\myclass{PP}}
\newcommand     {\NC}   {\myclass{NC}}
\newcommand     {\myAC}   {\myclass{AC}}
\newcommand     {\myACZ}   {\myAC^0}
\newcommand     {\myACZP}   {\myAC^0+\text{\myclass{Mod 2}}}
\newcommand     {\myTC}   {\myclass{TC}}
\newcommand     {\myTCZ}   {\myTC^0}

\newcommand     {\TC}   {\myclass{TC}}
\newcommand     {\SAC}   {\myclass{SAC}}
\newcommand     {\ACC}   {\myclass{ACC}}
\newcommand     {\tc}   {\myclass{TC}}   %
\newcommand     {\PPoly}{\myclass{\mbox{P}/\mbox{Poly}}} %

\newcommand     {\FOarb}   {\myclass{FO(arb)}}

\newcommand     {\NLIN}   {\myclass{NLIN}}
\newcommand     {\DLIN}   {\myclass{DLIN}}

\newcommand  {\APTIME}   {\myclass{APTIME}}
\newcommand  {\ALOGSPACE}   {\myclass{ALOGSPACE}}

\newcommand{\FO}{\StaClass{FO}}
\newcommand{\fotwo}{\StaClass{FO\ensuremath{^2}}}
\newcommand{\FOarithmetic}{\StaClass{FO$(+,\!\times\!)$}}
\newcommand{\MSO}[1][\quant]{\StaClass{MSO}}
\newcommand{\emsotwo}{\StaClass{$\exists$MSO$^2$}}
\newcommand{\EMSO}{\StaClass{$\exists$MSO}}
\newcommand{\ESO}{\StaClass{$\exists$SO}}
\newcommand{\QFO}[1][\quant]{\StaClass{\ensuremath{#1}FO}}
\newcommand{\cQFO}[1][\quant]{\StaClass{\ensuremath{\overline{#1}}FO}}
\newcommand{\EFO}{\QFO[\exists^*]}
\newcommand{\AFO}{\QFO[\forall^*]}
\newcommand{\AEFO}{\StaClass{$\forall/\exists$FO}}
\newcommand{\CQ}[1][]{\StaClass{CQ}}
\newcommand{\UCQ}[1][]{\StaClass{UCQ}}
\newcommand{\CQneg}[1][]{\StaClass{CQ\ensuremath{^{\mneg}}}}
\newcommand{\UCQneg}[1][]{\StaClass{UCQ\ensuremath{^{\mneg}}}}
\newcommand{\Prop}{\StaClass{Prop}}
\newcommand{\QF}{\StaClass{QF}}
\newcommand{\PropCQ}{\StaClass{PropCQ}}
\newcommand{\PropUCQ}{\StaClass{PropUCQ}}
\newcommand{\PropCQneg}{\StaClass{PropCQ{\ensuremath{^{\mneg}}}}}
\newcommand{\PropUCQneg}{\StaClass{PropUCQ{\ensuremath{^{\mneg}}}}}

\newcommand{\mneg}{\neg} %

\newcommand{\DynTC}{\DynClass{TC}}

\newcommand{\DynProp}{\DynClass{Prop}}
\newcommand{\DynPropbi}{\DynClass{Prop}^*\xspace}
\newcommand{\DynProj}{\DynClass{Projections}}
\newcommand{\DynQF}{\DynClass{QF}}
\newcommand{\DynQFbi}{\DynClass{QF}^*\xspace}
\newcommand{\DynFO}{\DynClass{FO}}
\newcommand{\DynFOfs}{\DynClass{FO}^\emptyset}
\newcommand{\DIDynFO}{\myclass{DI-DynFO}\xspace}
\newcommand{\DynFOpatnaik}{\myclass{Dyn-FO}\xspace}
\newcommand{\DynFOpatnaikstr}{\myclass{Dyn-FO$^+$}\xspace}
\newcommand{\DynFOarithmetic}{\DynClass{FO$(+,\!\times\!)$}\xspace}
\newcommand{\DynACZ}{\DynClass{AC$^0$}}
\newcommand{\DynACZP}{\DynClass{AC$^0 + $ Mod 2}}
\newcommand{\DynFOpos}{\DynClass{FO$^{\wedge, \vee}$}}
\newcommand{\DynFOand}{\DynClass{FO$^{\wedge}$}}

\newcommand{\DynC}{\DynClass{$\class$}}
\newcommand{\DynUCQ}{\DynClass{UCQ}}
\newcommand{\DynCQ}{\DynClass{CQ}}
\newcommand{\DynUCQneg}{\DynClass{UCQ$^\mneg$}}
\newcommand{\DynCQneg}{\DynClass{CQ$^\mneg$}}
\newcommand{\DynCQPM}{\DynCQneg}
\newcommand{\DyncQFO}{\DynClass{$\cquant$FO}}

\newcommand{\DynQFO}[1][\quant]{\DynClass{\QFO[#1]}}
\newcommand{\DynEFO}{\DynQFO[\exists^*]}
\newcommand{\DynAFO}{\DynQFO[\forall^*]}

\newcommand{\DynAEFO}{\DynClass{$\forall/\exists$FO}}
\newcommand{\DynAND}{\DynClass{PropCQ}}
\newcommand{\DynAnd}{\DynAND}
\newcommand{\DynPropCQ}{\DynAND}

\newcommand{\DynPropPos}{\DynClass{PropUCQ}}
\newcommand{\DynPropAO}{\DynPropPos}
\newcommand{\DynPropUCQ}{\DynPropPos}

\newcommand{\DynAndNeg}{\DynClass{PropCQ{\ensuremath{^{\mneg}}}}}
\newcommand{\DynPropCQneg}{\DynAndNeg}
\newcommand{\DynPropUCQneg}{\DynClass{PropUCQ{\ensuremath{^{\mneg}}}}}

\newcommand{\DynOrNeg}{\DynClass{Or{\ensuremath{^{\mneg}}}}}

\newcommand{\dDynProp}{\dDynClass{Prop}}
\newcommand{\dDynPropPos}{\dDynClass{PropUCQ}}
\newcommand{\dDynAndOr}{\dDynPropPos}
\newcommand{\dDynQF}{\dDynClass{QF}}
\newcommand{\dDynFO}{\dDynClass{FO}}
\newcommand{\dDynFOpos}{\dDynClass{FO$^{\wedge, \vee}$}}
\newcommand{\dDynFOand}{\dDynClass{FO$^{\wedge}$}}
\newcommand{\dDynC}{\dDynClass{$\class$}}
\newcommand{\dDynUCQ}{\dDynClass{UCQ}}
\newcommand{\dDynCQ}{\dDynClass{CQ}}
\newcommand{\dDynUCQneg}{\dDynClass{UCQ$^\mneg$}}
\newcommand{\dDynCQneg}{\dDynClass{CQ$^\mneg$}}
\newcommand{\dDynCQPM}{\dDynCQneg}

\newcommand{\dDynQFO}[1][\quant]{\dDynClass{\QFO[#1]}}
\newcommand{\dDynEFO}{\dDynQFO[\exists^*]}
\newcommand{\dDynAFO}{\dDynQFO[\forall^*]}

\newcommand{\dDynAEFO}{\dDynClass{$\forall/\exists$FO}}
\newcommand{\dDynAND}{\dDynPropCQ}
\newcommand{\dDynAnd}{\dDynAND}
\newcommand{\dDynConj}{\dDynClass{Conj}}
\newcommand{\dDynPropAO}{\dDynClass{Prop$^{\wedge, \vee}$}}
\newcommand{\dDyncQFO}{\dDynClass{$\cquant$FO}}
\newcommand{\dDynPropUCQneg}{\dDynClass{PropUCQ{\ensuremath{^{\mneg}}}}}
\newcommand{\dDynPropUCQ}{\dDynClass{PropUCQ}}
\newcommand{\dDynPropCQneg}{\dDynClass{PropCQ{\ensuremath{^{\mneg}}}}}
\newcommand{\dDynPropCQ}{\dDynClass{PropCQ}}

\newcommand{\equalcardinality}{\textsc{EqualCardinality}\xspace}%
\newcommand{\reach}{\textsc{Reach}\xspace}%
\newcommand{\altreach}{\textsc{Alt-Reach}\xspace}%

\newcommand{\stgraph}{$s$-$t$-graph\xspace}
\newcommand{\stgraphs}{$s$-$t$-graphs\xspace}
\newcommand{\reachQ}{\textsc{Reach}\xspace}
\newcommand{\streachQ}{\textsc{$s$-$t$-Reach}\xspace}
\newcommand{\streachabilityquery}{$s$-$t$-reachability query\xspace}
\newcommand{\stTwoPath}{\problem{$s$-$t$-Two\-Path}\xspace}
\newcommand{\sTwoPath}{\problem{$s$-Two\-Path}\xspace}
\makeatletter
\newcommand{\clique}[1]{%
  \@ifmtarg{#1}{%
    \problem{Clique}\xspace%
  }{%
    \problem{$#1$-Clique}\xspace%
  }%
}
\newcommand{\colorability}[1]{\problem{$#1$-Col}\xspace}
\newcommand{\streach}{$s$-$t$-Reach}
\newcommand{\streachp}{\problem{\streach}\xspace}
\newcommand{\layeredstreach}[1]{#1-Layered-$s$-$t$-Reach}
\newcommand{\layeredstreachp}[1]{\problem{\layeredstreach{#1}}\xspace}

\newcommand{\kmod}[2]{\textsc{$(k,#2)$-Mod-#1}}

\newcommand{\dynClique}[1]{\clique{#1}}
\newcommand{\dynColorability}[1]{\colorability{#1}}

\newcommand{\probEqualCardinalityText}{EqualCardinality}
\newcommand{\EqualCardinality}{\problem{\probEqualCardinalityText}\xspace}
\newcommand{\EqualCardinalityDescr}{\problemdescr{\probEqualCardinalityText}{Unary relations $A$ and $B$}{Do $A$ and $B$ have the same cardinality?\xspace}}

\newcommand{\dynEqualCardinality}{\dynProb{\probEqualCardinalityText}\xspace}
\newcommand{\dynEqualCardinalityDescr}{\dynProbDescr{\probEqualCardinalityText}{Unary relations $A$ and $B$}{Element insertions and deletions}{Do $A$ and $B$ have the same cardinality?\xspace}}

\newcommand{\dynReachQ}{\dynProb{\textsc{Reach}}\xspace}
\newcommand{\dynstReachQ}{\dynProb{\textsc{$s$-$t$-Reach}}\xspace}

\newcommand{\dynstTwoPath}{\dynProb{\stTwoPath}\xspace}
\newcommand{\dynsTwoPath}{\dynProb{\sTwoPath}\xspace}

\newcommand{\dynlayeredstreach}[1]{Dyn-#1-Layered-$s$-$t$-Reach}
\newcommand{\dynlayeredstreachp}[1]{\problem{\dynlayeredstreach{#1}}\xspace}

\makeatletter{}%
\newcommand{\mtext}[1]{\textsc{#1}}

\providecommand {\calA}      {{\mathcal A}\xspace}
\providecommand {\calB}      {{\mathcal B}\xspace}
\providecommand {\calC}      {{\mathcal C}\xspace}
\providecommand {\calD}      {{\mathcal D}\xspace}
\providecommand {\calE}      {{\mathcal E}\xspace}
\providecommand {\calF}      {{\mathcal F}\xspace}
\providecommand {\calG}      {{\mathcal G}\xspace}
\providecommand {\calH}      {{\mathcal H}\xspace}
\providecommand {\calK}      {{\mathcal K}\xspace}
\providecommand {\calI}      {{\mathcal I}\xspace}
\providecommand {\calL}      {{\mathcal L}\xspace}
\providecommand {\calM}      {{\mathcal M}\xspace}
\providecommand {\calN}      {{\mathcal N}\xspace}
\providecommand {\calO}      {{\mathcal O}\xspace}
\providecommand {\calP}      {{\mathcal P}\xspace}
\providecommand {\calQ}      {{\mathcal Q}\xspace}
\providecommand {\calR}      {{\mathcal R}\xspace}
\providecommand {\calS}      {{\mathcal S}\xspace}
\providecommand {\calT}      {{\mathcal T}\xspace}
\providecommand {\calU}      {{\mathcal U}\xspace}
\providecommand {\calV}      {{\mathcal V}\xspace}
\providecommand {\calX}      {{\mathcal X}\xspace}
\providecommand {\calZ}      {{\mathcal Z}\xspace}

\newcommand{\mhat}[1]{\widehat{#1}}

\newcommand{\Ah}{\mhat{A}}
\newcommand{\Bh}{\mhat{B}}
\newcommand{\Ch}{\mhat{C}}
\newcommand{\Dh}{\mhat{D}}
\newcommand{\Eh}{\mhat{E}}
\newcommand{\Fh}{\mhat{F}}
\newcommand{\Gh}{\mhat{G}}
\newcommand{\Hh}{\mhat{H}}
\newcommand{\Ih}{\mhat{I}}
\newcommand{\Jh}{\mhat{J}}
\newcommand{\Kh}{\mhat{K}}
\newcommand{\Lh}{\mhat{L}}
\newcommand{\Mh}{\mhat{M}}
\newcommand{\Nh}{\mhat{N}}
\newcommand{\Oh}{\mhat{O}}
\newcommand{\Ph}{\mhat{P}}
\newcommand{\Qh}{\mhat{Q}}
\newcommand{\Rh}{\mhat{R}}
\newcommand{\Sh}{\mhat{S}}
\newcommand{\Th}{\mhat{T}}
\newcommand{\Uh}{\mhat{U}}
\newcommand{\Vh}{\mhat{V}}
\newcommand{\Wh}{\mhat{W}}
\newcommand{\Xh}{\mhat{X}}
\newcommand{\Yh}{\mhat{Y}}
\newcommand{\Zh}{\mhat{Z}}

\newcommand{\Psih}{\mhat{\Psi}}
\newcommand{\psih}{\mhat{\psi}}
\newcommand{\Phih}{\mhat{\Phi}}
\newcommand{\phih}{\mhat{\phi}}
\newcommand{\varphih}{\mhat{\varphi}}

\newcommand{\eqh}{\mhat{=}}

\newcommand{\Z}{\ensuremath{\mathbb{Z}}}

\newcommand{\N}{\ensuremath{\mathbb{N}}}

\newcommand{\Q}{\ensuremath{\mathbb{Q}}}

\newcommand{\R}{\ensuremath{\mathbb{R}}}

\newcommand{\perm}{\ensuremath{\pi}}

\newcommand{\allsubsets}[2]{[#1]^{#2}}
\newcommand{\pvec}[1]{\vec{#1}\mkern2mu\vphantom{#1}}

\newcommand{\kexp}[2]{\ensuremath{\exp^{#1}\hspace{-0.5mm}(#2)}}

\newcommand{\tower}[2]{\ensuremath{\text{tow}_{#1}\hspace{-0.5mm}(#2)}}

\newcommand{\klog}[2]{\ensuremath{\log^{(#1)}{\hspace{-0.5mm}(#2)}}}
\newcommand{\klogwb}[2]{\ensuremath{\log^{(#1)}{\hspace{-0.5mm}#2}}}

\newcommand{\disjointunion}{\uplus}

\providecommand{\power}[1]{\ensuremath{\calP(#1)}\xspace}

\newcommand{\restrict}[2]{#1\mspace{-3mu}\upharpoonright \mspace{-3mu}#2}

\newcommand{\isomorph}{\simeq}
\newcommand{\isomorphVia}[1]{\isomorph_{#1}}
\newcommand{\swap}[2]{id{[#1, #2]}}

\newcommand{\df}{\ensuremath{\mathrel{\smash{\stackrel{\scriptscriptstyle{
    \text{def}}}{=}}}} \;}

\newcommand{\refeq}[1]{\ensuremath{{\stackrel{\scriptstyle{
    \text{#1}}}{=}}}}

\newcommand{\longlongeq}{=\joinrel=\joinrel=\joinrel=}
\newcommand{\longeq}{=\joinrel=\joinrel=}
\newcommand{\reflongeq}[1]{\ensuremath{{\stackrel{\scriptstyle{
    \text{#1}}}{\longeq}}}}

\newcommand{\ramseyw}[1]{\ensuremath{R_{#1}}}

\makeatletter %
\newcommand{\auxramsey}[4]{
  \@ifmtarg{#1}{
    \@ifmtarg{#4}{
      \ensuremath{R(#2; #3)}
    }{
      \ensuremath{R^#4(#2; #3)}
    }
   }{
    \@ifmtarg{#4}{
      \ensuremath{R_{#1}(#2; #3)}
    }{
      \ensuremath{R^#4_{#1}(#2; #3)}
    }
  }
}

\newcommand{\ramsey}[3]{\auxramsey{#1}{#2}{#3}{}}
\newcommand{\homramsey}[2]{\auxramsey{}{#2}{#1}{\text{hom}}}
\newcommand{\mfoldramsey}[3]{\auxramsey{}{#2}{#1}{#3}}

\newcommand{\norder}{\prec}

\newcommand{\col}{col}

\newcommand{\property}{($\ast$)}

\newcommand{\subseq}{\sqsubseteq}

\newcommand{\derive}{\Rightarrow}
\newcommand{\rmapsto}{\rightarrow}

\newcommand{\lpath}[1][]{\ensuremath{\mathrel{\smash{\stackrel{\scriptscriptstyle{
    #1}}{\rightsquigarrow}}}}}

\makeatletter{}%

   \theoremstyle{plain}

   \newtheorem{proposition}[theorem]{Proposition}
   \newtheorem{fact}[theorem]{Fact}
   \newtheorem*{claim*}{Claim}
   \newtheorem*{fact*}{Fact}
   \newtheorem{claim}{Claim}

    \newtheorem{goal}{Goal}
    \newtheorem*{goal*}{Goal}
    \theoremstyle{definition}

    \newtheorem {openquestion}{Open question}
    \newtheorem {openproblem}{Open problem}
    \newtheorem {question}{Question}
    \newtheorem {mainquestion}{Main question}

    \newenvironment{proofsketch}{\begin{proof}[Proof sketch.]}{\end{proof}}
    \newenvironment{proofidea}{\begin{proof}[Proof idea.]}{\end{proof}}

    \newenvironment{proofof}[1]{\begin{proof}[Proof (of #1).]}{\end{proof}}
   \newenvironment{proofsketchof}[1]{\begin{proof}[Proof sketch (of #1).]}{\end{proof}}
   \newenvironment{proofideaof}[1]{\begin{proof}[Proof idea (of #1).]}{\end{proof}}

\newenvironment{proofenum}{\begin{enumerate}[label=(\alph*),wide=0pt, listparindent=15pt]}{\end{enumerate}}

\makeatletter{}%
\newcommand{\eval}[3]{#1(#2/#3)}

\newcommand{\assignment}{\theta}

\newcommand{\arity}{\ensuremath{\text{Ar}}}
\newcommand{\arityFun}{\ensuremath{Ar_{\text{fun}}}}

\newcommand{\schema}{\tau}
\newcommand{\schemah}{\mhat{\schema}}
\newcommand{\relSchema}{\schema_{\text{rel}}}
\newcommand{\relSchemah}{\schemah_{\text{rel}}}
\newcommand{\conSchema}{\schema_{\text{const}}}
\newcommand{\conSchemah}{\schemah_{\text{const}}}
\newcommand{\funSchema}{\schema_{\text{fun}}}
\newcommand{\funSchemah}{\schemah_{\text{fun}}}
\newcommand{\Terms}[2]{\textsc{Terms}^{#2}_{#1}} 

\newcommand{\struc}{\calS}
\newcommand{\struca}{\struc}
\newcommand{\strucb}{\calT}

\newcommand{\unaryTypes}[1]{\mathcal{UN}_{#1}}
\newcommand{\binaryTypes}[1]{\mathcal{BIN}_{#1}}
\newcommand{\naryTypes}[2]{\mathfrak{T}_{#1,#2}}

\newcommand{\nb}[3]{\calN_{#2}^{#3}(#1)}
\newcommand{\nbv}[3]{\vec \calN_{#2}^{#3}(#1)}

\newcommand{\mthen}{\rightarrow}
\newcommand{\mand}{\wedge}
\newcommand{\mor}{\vee}
\newcommand{\munion}{\cup}
\newcommand{\mintersect}{\cap}
\newcommand{\mdisjunion}{\biguplus}
\newcommand{\sem}[2]{\ensuremath{\llbracket #1\rrbracket_{#2}}} %

\newcommand{\arb}{\ensuremath{\star}}%
\newcommand{\generic}{\textsc{generic}}
\newcommand{\quant}{\mathbb{Q}}
\newcommand{\cquant}{\overline{\mathbb{Q}}}

\newcommand{\nd}{d}
\newcommand{\formulas}{\calC}
\newcommand{\symneg}[1]{\widehat{#1}}

\newcommand{\type}[2]{\ensuremath{\langle #1, #2 \rangle}}
\newcommand{\stype}[3]{\ensuremath{\langle #1, #2 \rangle_{#3}}}

\newcommand{\behaveEqual}[1]{\approx_{#1}}

\newcommand{\types}[2]{types_{#1}(#2)}
\newcommand{\numTypes}[2]{|\types{{#1}}{#2}|}

\newcommand{\eqtype}{\epsilon}

\newcommand{\db}{\calD}
\newcommand{\inp}{\calI}
\newcommand{\aux}{\calA}
\newcommand{\builtin}{\calB}
\newcommand{\domain}{D}
\newcommand{\actDomain}{D_\mtext{act}}

\newcommand{\query}{\calQ}
\newcommand{\cq}{\calC}

\newcommand{\querys}{Q}

\newcommand{\ans}[2]{\mtext{ans}(#1, #2)}

\newcommand{\updates}{\ensuremath{\Delta}}
\newcommand{\abstrDel}{\ensuremath{\updates_{Del}}}
\newcommand{\abstrIns}{\ensuremath{\updates_{Ins}}}
\newcommand{\abstrUpd}{\ensuremath{\updates}}

\newcommand{\init}{\mtext{Init}\xspace}

\newcommand{\ins}{\mtext{ins\xspace}}
\newcommand{\del}{\mtext{del\xspace}}

\newcommand{\insertdescr}[2]{\textbf{Insertion of \ensuremath{#2} into \ensuremath{#1}.}}
\newcommand{\deletedescr}[2]{\textbf{Deletion of \ensuremath{#2} from \ensuremath{#1}.}}

\newcommand{\state}{\struc}
\newcommand{\advDB}{\db_{\text{help}}}
\newcommand{\advState}{\state_{\text{help}}}

\newcommand{\inpSchema}{\schema_{\text{in}}}
\newcommand{\auxSchema}{\schema_{\text{aux}}}
\newcommand{\advSchema}{\schema_{\text{help}}}
\newcommand{\eqSchema}{\schema_{=}}
\newcommand{\builtinSchema}{\schema_{\text{bi}}}

\newcommand{\auxInit}{\init_{\text{aux}}}
\newcommand{\builtinInit}{\init_{\text{bi}}}

\providecommand{\prog}{\ensuremath{\calP}\xspace}
\newcommand{\progb}{\ensuremath{Q}\xspace}

\newcommand{\updateDB}[2]{\ensuremath{#1(#2)}}
\newcommand{\updateState}[3]{\ensuremath{#1_{#2}(#3)}}
\newcommand{\updateRelation}[4]{\restrict{\ensuremath{{#1}_{#2}(#3)}}{#4}}

\newcommand{\transition}[3]{\ensuremath{{#1} \xrightarrow{#2}{#3}}}

\makeatletter %
\newcommand{\uf}[4]{
  \@ifmtarg{#4}{
    \ensuremath{\phi^{#1}_{#2}(#3)}
   }{
    \ensuremath{\phi^{#1}_{#2}(#3; #4)}
  }
}
\newcommand{\huf}[4]{
  \@ifmtarg{#4}{
    \ensuremath{\widehat{\phi}^{#1}_{#2}(#3)}
   }{
    \ensuremath{\widehat{\phi}^{#1}_{#2}(#3; #4)}
  }
}

\newcommand{\ufb}[4]{
  \@ifmtarg{#4}{
    \ensuremath{\psi^{#1}_{#2}(#3)}
   }{
    \ensuremath{\psi^{#1}_{#2}(#3; #4)}
  }
}

\newcommand{\ufbwa}[2]{
  \ensuremath{\psi^{#1}_{#2}}
}

\newcommand{\ufwa}[2]{
  \ensuremath{\phi^{#1}_{#2}}
}

  \makeatletter %
  \newcommand{\ufsubstitute}[5]{
    \@ifmtarg{#5}{
      \ensuremath{\phi^{#2}_{#3}[#1](#4)}
    }{
      \ensuremath{\phi^{#2}_{#3}[#1](#4; #5)}
    }
  }

  \makeatletter %
  \newcommand{\ufsubstitutewa}[3]{
      \ensuremath{\phi^{#2}_{#3}[#1]}
  }
  \makeatletter %
  \newcommand{\substitutewa}[2]{
      \ensuremath{#1[#2]}
  }

\newcommand{\ut}[4]{
  \@ifmtarg{#4}{
    \ensuremath{t^{#1}_{#2}(#3)}
   }{
    \ensuremath{t^{#1}_{#2}(#3; #4)}
  }
}

\newcommand{\utw}[3]{
  \ensuremath{t^{#1}_{#2}(#3)}
}

\newcommand{\utwa}[2]{\ensuremath{t^{#1}_{#2}}}
\newcommand{\ITE}{\mtext{ITE}}
\newcommand{\ite}[3]{
  \@ifmtarg{#1}{
    \ensuremath{\ITE}
   }{
    \ITE(#1,#2,#3)  
  }
}

\newcommand{\itewa}{
    \ensuremath{\ITE}
}
\makeatletter{}%

\providecommand{\nc}{\newcommand}
\providecommand{\rnc}{\renewcommand}
\providecommand{\pc}{\providecommand}

\renewcommand{\labelenumi}{(\alph{enumi})}

\newcommand{\Erdos}{Erd\H{o}s}

\newcommand{\quotes}[1]{``#1''}

\newcommand{\mbold}[1]{\textbf{\hyperpage{#1}}}
\newcommand{\mitalic}[1]{{\it #1}}

\newcommand{\bfindex}[1]{\index{#1|mbold}}
\newcommand{\itindex}[1]{\index{#1|it}}

\newcommand{\defindex}[1]{\bfindex{#1}}

\ifcomments
\nc{\commentbox}[1]{\noindent\framebox{\parbox{\linewidth}{#1}}}
\nc{\todo}[1]{\ \\ {\color{red} \fbox{\parbox{\linewidth}{{\sc
          ToDo}:\\  #1}}}}

\setlength{\marginparwidth}{2.5cm}
\setlength{\marginparsep}{3pt}

\newcounter{CommentCounter}
\newcommand{\acomment}[2]{\ \\ \fbox{\parbox{\linewidth}{{\sc #1}: #2}}}
\newcommand{\mcomment}[2]{{\color{blue}(#1)}\footnote{#1: #2}} %
\else
\nc{\commentbox}[1]{}
\newcommand{\mcomment}[2]{}
\newcommand{\acomment}[2]{}
\fi

\ifchanges

\newcommand{\loldnew}[3]{\commentbox{{\textcolor{blue}{\setlength{\fboxsep}{1pt}\fbox{\small
          #1}}} \textcolor{red}{\footnotesize #2}}
  \textcolor{blue}{#3}}
\setul{}{0.2mm}
\setstcolor{red}
\newcommand{\oldnew}[3]{{\textcolor{blue}{\setlength{\fboxsep}{1pt}\fbox{\small
        #1}}} \st{\footnotesize #2} \textcolor{blue}{#3}}

\else
\newcommand{\loldnew}[3]{#3}
\newcommand{\oldnew}[3]{#3}
\fi

\nc{\tzm}[1]{\mcomment{TZ}{#1}}
\nc{\tsm}[1]{\mcomment{TS}{#1}}
\nc{\pablom}[1]{\mcomment{PM}{#1}}
\nc{\nilsm}[1]{\mcomment{NV}{#1}}
\nc{\tz}[1]{\acomment{TZ}{#1}}
\nc{\thz}[1]{\acomment{TZ}{#1}}
\nc{\ts}[1]{\acomment{TS}{#1}}
\nc{\pablo}[1]{\acomment{PM}{#1}}
\nc{\nils}[1]{\acomment{NV}{#1}}

\nc{\tzon}[2][]{\oldnew{TZ}{#1}{#2}} 
\nc{\tson}[2][]{\oldnew{TS}{#1}{#2}}

\nc{\tzlon}[2][]{\loldnew{TZ}{#1}{#2}} 
\nc{\tslon}[2][]{\loldnew{TS}{#1}{#2}}

\makeatletter{}%
\newcommand{\apptheoremtitlefont}[1]{\textbf{#1}}
\newcommand{\apptheoremcontentfont}{\itshape}

\newcommand{\apponlystartmarker}{ $\blacktriangleright\blacktriangleright\blacktriangleright$ }
\newcommand{\apponlyendmarker}{ $\blacktriangleleft\blacktriangleleft\blacktriangleleft$ }
\newcommand{\apprepetitionstartmarker}{ $\blacktriangleright\blacktriangleright\blacktriangleright$ }
\newcommand{\apprepetitionendmarker}{ $\blacktriangleleft\blacktriangleleft\blacktriangleleft$ }

\newcommand{\initialAppendix}{
  \section*{Appendix}

  In the appendix we give the proofs that have been omitted in the main text. For proofs that are partially present in the main article, we repeat the full proof and its context. Parts that are only repeated are marked by \apprepetitionstartmarker and \apprepetitionendmarker. 
  
}
  
\newcommand{\writeAppendix}{
  
  \initialAppendix
}
\newcommand{\toAppendix}[1]{
  \makeatletter
   \g@addto@macro\writeAppendix{#1}
  \makeatother
}

\newcommand{\toMainAndAppendix}[1]{
  \longVersion{#1}
  \shortVersion{
    #1%
    \toAppendix{%
	\apprepetition{#1} \par
    }
  }
}

\newcommand{\toLongAndAppendix}[1]{
  \longVersion{#1}%
  \shortVersion{    
    \toAppendix{%
	#1 \par
    }
  }
}

\newcommand{\atheorem}[2]{
  \begin{theorem}\label{#1}%
    #2%
  \end{theorem}%
  \toAppendix{%
    \begin{apptheorem}{\ref{#1}}{}
      #2%
    \end{apptheorem}%
  }
}

\newcommand{\alemma}[2]{
  \begin{lemma}\label{#1}%
    #2%
  \end{lemma}%
  \toAppendix{%
    \begin{applemma}{\ref{#1}}{}%
      #2%
    \end{applemma}%
  }
}

\newcommand{\aproposition}[2]{
  \begin{proposition}\label{#1}%
    #2%
  \end{proposition}%
  \toAppendix{%
    \begin{appproposition}{\ref{#1}}{}%
      #2%
    \end{appproposition}%
  }
}

\newcommand{\aproof}[3]{%
  \longVersion{
    \begin{proof}%
      #1%
      #3%
    \end{proof}%
  }
  \shortVersion{
      \begin{proof}%
	#1%
	#2%
      \end{proof}%
    \toAppendix{%
      \begin{proof}%
	\@ifmtarg{#1}{}{\apprepetition{#1}} \par
	#3
      \end{proof}%
    }
  }
}

\newcommand{\aproofof}[4]{%
  \longVersion{
    \begin{proofof}{#1}%
      #2%
      #4%
    \end{proofof}%
  }
  \shortVersion{
      \begin{proofof}{#1}%
	#2%
	#3%
      \end{proofof}%
    \toAppendix{%
      \begin{proof}%
	\@ifmtarg{#2}{}{\apprepetition{#2}} \par
	#4
      \end{proof}%
    }
  }
}

\newcommand{\aproofsketch}[3]{%
  \longVersion{
    \begin{proof}%
      #1%
      #3%
    \end{proof}%
  }

  \shortVersion{
    \@ifmtarg{#2}{}{%
      \begin{proofsketch}%
	#1%
	#2%
      \end{proofsketch}%
    }
    \toAppendix{%
      \begin{proof}%
	\@ifmtarg{#1}{}{\apprepetition{#1}} \par
	#3
      \end{proof}%
    }
  }
}

\newcommand{\aproofsketchidea}[3]{%
  \longVersion{
    \begin{proofsketch}%
      #1%
      #3%
    \end{proofsketch}%
  }

  \shortVersion{
    \@ifmtarg{#2}{}{%
      \begin{proofidea}%
	#1%
	#2%
      \end{proofidea}%
    }
    \toAppendix{%
      \begin{proofsketch}%
	\@ifmtarg{#1}{}{\apprepetition{#1}} \par
	#3
      \end{proofsketch}%
    }
  }
}

\newcommand{\aproofsketchideaof}[4]{%
  \longVersion{
    \begin{proofsketchof}{#1}%
      #2%
      #4%
    \end{proofsketchof}%
  }

  \shortVersion{
    \@ifmtarg{#2}{}{%
      \begin{proofideaof}{#1}%
	#2%
	#3%
      \end{proofideaof}%
    }
    \toAppendix{%
      \begin{proofsketchof}{#1}%
	\@ifmtarg{#3}{}{\apprepetition{#3}} \par
	#4
      \end{proofsketchof}%
    }
  }
}

\newcommand{\aproofidea}[3]{%
  \longVersion{
    \begin{proof}%
      #1%
      #3%
    \end{proof}%
  }

  \shortVersion{
    \@ifmtarg{#2}{}{%
      \begin{proofidea}%
	#1%
	#2%
      \end{proofidea}%
    }
    \toAppendix{%
      \begin{proof}%
	\@ifmtarg{#1}{}{\apprepetition{#1}} \par
	#3
      \end{proof}%
    }
  }
}

\newcommand{\shortOrLong}[2]{%
  \shortVersion{#1}%
  \longVersion{#2}%
}

\makeatletter
\newcommand{\theoremcont}[3]{
   \def\Type{#1}
   \def\Number{#2}
   \def\Label{#3}
  \@ifmtarg{#3}{
     \apptheoremtitlefont{\Type\ \Number.} \apptheoremcontentfont
   }{
    \apptheoremtitlefont{\Type\ \Number}\ \apptheoremcontentfont(\Label).
  }
}

\newenvironment{applemma}[2]{\vspace{2mm}\par\theoremcont{Lemma}{#1}{#2}}{\vspace{0mm}\par}
\newenvironment{apptheorem}[2]{\vspace{2mm}\par\theoremcont{Theorem}{#1}{#2}}{\vspace{2mm} \par }
\newenvironment{appcorollary}[2]{\theoremcont{Corollary}{#1}{#2}}{\vspace{2mm}}
\newenvironment{appproposition}[2]{\vspace{2mm}\theoremcont{Proposition}{#1}{#2}}{\vspace{2mm} \par}
\newenvironment{appdefinition}[2]{\theoremcont{Definition}{#1}{#2}}{\vspace{2mm}}
\newenvironment{appexample}[1]{\vspace{2mm}\textit{Example #1.}}{\vspace{2mm}}

\newcommand{\apponlystart}{
  \apponlystartmarker
}
\newcommand{\apponlyend}{
  \apponlyendmarker
}

\newcommand{\apprepetition}[1]{
  \apprepetitionstartmarker #1 \apprepetitionendmarker
}

\newcommand{\LineIf}[2]{\State \algorithmicif\ {#1}\ \algorithmicthen\ {#2}}

\newcommand{\columnWidth}{11cm}

\newcommand{\substruclemma}{Substructure Lemma\xspace}
\newcommand{\First}{\mtext{First}}
\newcommand{\List}{\mtext{List}}
\newcommand{\Last}{\mtext{Last}}
\newcommand{\In}{\mtext{In}}
\newcommand{\Out}{\mtext{Out}}
\newcommand{\Empty}{\mtext{Empty}}

\newcommand{\Odd}{\mtext{Odd}}
\newcommand{\odd}{\text{odd}}
\newcommand{\even}{\text{even}}

\newcommand{\Counter}{\mtext{Cnt}}
\newcommand{\isEmpty}{\mtext{Empty}}
\newcommand{\Zero}{\mtext{Zero}}

\newcommand{\congruent}[2]{\sim_{#1, #2}}

\newcommand{\Succ}{\mtext{Succ}}
\newcommand{\Pred}{\mtext{Pred}}
\newcommand{\Max}{\mtext{Max}}
\newcommand{\numEdges}{\#\mtext{edges}}
\newcommand{\numNodes}{\#\mtext{nodes}}

\newcommand{\NRC}{\ensuremath{\mathcal{NRC}}\xspace}
\newcommand{\SQL}{\ensuremath{\mathcal{SQL}}\xspace}

\newcommand{\progToGraph}[1]{\ensuremath{{\langle #1 \rangle}}}

\algnewcommand\algorithmicinput{\textbf{Input:}}
\algnewcommand\INPUT{\item[\algorithmicinput]}

\algnewcommand\algorithmicoutput{\textbf{Output:}}
\algnewcommand\OUTPUT{\item[\algorithmicoutput]}

\newcommand{\C}{\mathcal C}
\newcommand{\prodreach}{$\times$-\reach}
\renewcommand{\L}{\mathcal L}
\newcommand{\nlog}{\textsc{NLogspace}}
\newcommand{\np}{\textsc{NP}}
\newcommand{\pspace}{\textsc{PSpace}}

\newcommand{\NN}{\mathbb N}

\nc{\apad}{\mbox{\sc AllDist}\xspace}
\nc{\eqlength}{\mbox{\sc Equal-Length-Paths}\xspace}
\newcommand{\Appendix}{\centerline{------------ Material for Appendix ---------}}

\newcommand{\neps}{\textsc{Neps}\xspace}
\newcommand{\FF}{\mathbb F}
\newcommand{\rank}{\text{rank}}
\newcommand{\<}{\left <}
\renewcommand{\>}{\right >}
\newcommand{\DynTCZ}{{\sc DynTC$^0$}}

%
%
%
%
%
%
%
%
%
%
%
%
%
%
%
%
%
%

   %
\makeatletter{}%

\usetikzlibrary{arrows, shapes, snakes}

\tikzstyle{mnode}=[
  circle,
  fill=\mnodefillcolor, 
  draw=\mnodedrawcolor,
  minimum size=6pt, 
  inner sep=0pt
]

\tikzstyle{mnodeinvisible}=[
  minimum size=6pt, 
  inner sep=0pt
]

\tikzstyle{invisible}=[
  minimum size=0pt, 
  inner sep=0pt
]

\tikzstyle{invisiblel}=[
  minimum size=10pt, 
  inner sep=0pt
]

\tikzstyle{invisibleEdge}=[
  transparent
]

\tikzstyle{nameNode}=[
  font=\scriptsize
]

\tikzstyle{namingNode}=[
  font=\normalsize
]

\tikzstyle{mEdge}=[
  -latex', %
  thick, 
  shorten >=3pt, 
  shorten <=3pt,
  draw=black!80,
]

\tikzstyle{dDashedEdge}=[
  -latex', %
  thick, 
  shorten >=3pt, 
  shorten <=3pt,
  draw=black!80,
  dashed
]

\tikzstyle{dEdge}=[
  -latex', %
  thick, 
  shorten >=3pt, 
  shorten <=3pt,
  draw=black!80,
]

\tikzstyle{dhEdge}=[
  -latex', %
  thick, 
  shorten >=3pt, 
  shorten <=3pt,
  draw=black!80,
]
\tikzstyle{uEdge}=[
  thick, 
  shorten >=3pt, 
  shorten <=3pt,
  draw=black!80,
]
\tikzstyle{uhEdge}=[
  thick, 
  shorten >=3pt, 
  shorten <=3pt,
  draw=black!80,
]

\tikzstyle{cEdge}=[
  ultra thick, 
  shorten >=3pt, 
  shorten <=3pt,
  draw=black!80,
]

\tikzstyle{dotsEdge}=[
  very thick, 
  loosely dotted, 
  shorten >=7pt, 
  shorten <=7pt
]

\tikzstyle{snakeEdge}=[
  ->, 
  decorate, 
  decoration={snake,amplitude=.4mm,segment length=2.5mm,post length=0.5mm},
]

\tikzstyle{snakeEdgea}=[
  ->, 
  decorate, 
  decoration={snake,amplitude=.4mm,segment length=3mm,post length=0.5mm}
]

\newcommand{\redEdge}{\tikz{
  \node (tmpa) at (-0.4,0)[mnode]{};
  \node (tmpb) at (0.4,0)[mnode]{};
  \draw [cEdge, draw=red] (tmpa) to (tmpb);
}}

\newcommand{\yellowEdge}{\tikz{
  \node (tmpa) at (-0.4,0)[mnode]{};
  \node (tmpb) at (0.4,0)[mnode]{};
  \draw [cEdge, draw=yellow] (tmpa) to (tmpb);
}}

\newcommand{\blueEdge}{\tikz{
  \node (tmpa) at (-0.4,0)[mnode]{};
  \node (tmpb) at (0.4,0)[mnode]{};
  \draw [cEdge, draw=blue] (tmpa) to (tmpb);
}}

\newcommand{\violetEdge}{\tikz{
  \node (tmpa) at (-0.4,0)[mnode]{};
  \node (tmpb) at (0.4,0)[mnode]{};
  \draw [cEdge, draw=violet] (tmpa) to (tmpb);
}}

\newcommand{\textnode}[1]{%
      \begin{minipage}{100pt}%
    #1
      \end{minipage}%
}

\newcommand{\classNode}[2]{
    \begin{scope}[shift={#1}]
      \node[text depth=0.7cm, anchor=north] (tmp) at (0,0) {
  \begin{minipage}{170pt}%
    #2
  \end{minipage}%
      };
    \end{scope}
}

\tikzstyle{class rectangle}=[
  draw=black,
  inner sep=0.2cm,
  rounded corners=5pt,
  thick
]

\tikzstyle{mline}=[
  draw=black,
  inner sep=0.2cm,
  rounded corners=5pt,
  thick
]

\tikzstyle{mainclass rectangle}=[
  draw=blue,
  inner sep=0.2cm,
  rounded corners=5pt,
  very thick
]

\newcommand{\redpoint}{\tikz{\node (tmp) at (-1.5,1.6)[mnode, fill=red]{};}}
\newcommand{\yellowpoint}{\tikz{\node (tmp) at (-1.5,1.6)[mnode, fill=yellow]{};}}
\newcommand{\bluepoint}{\tikz{\node (tmp) at (-1.5,1.6)[mnode, fill=blue]{};}}
\newcommand{\violetpoint}{\tikz{\node (tmp) at (-1.5,1.6)[mnode, fill=violet]{};}}

\newcommand{\bgcolor}{black!5}
\newcommand{\substructurefillcolor}{blue!40}
\newcommand{\substructureufillcolor}{blue!60!red!40}
\newcommand{\substructuredrawcolor}{blue!80}
\newcommand{\substructureudrawcolor}{blue!60!red!80}

\newcommand{\structurefillcolor}{blue!20}
\newcommand{\structureufillcolor}{blue!60!red!20}
\newcommand{\structuredrawcolor}{blue!40}
\newcommand{\structureudrawcolor}{blue!60!red!40}

\newcommand{\mnodedrawcolor}{black!80}
\newcommand{\mnodefillcolor}{black!40}

\newcommand{\mnc}{\mnodefillcolor}
\newcommand{\mgrey}{\mnodefillcolor}
\newcommand{\mblue}{blue}
\newcommand{\mred}{red}
\newcommand{\myellow}{yellow}
\newcommand{\mviolet}{violet}

\newcommand{\mopacity}{1.0}
\newcommand{\opac}{0.2}

\pgfdeclarelayer{background}
\pgfdeclarelayer{substructure}
\pgfdeclarelayer{edges}
\pgfdeclarelayer{foreground}
\pgfsetlayers{background,substructure,edges,main,foreground}

\tikzstyle{background rectangle}=[
  fill=black!10,
  draw=black!20,
  line width = 5pt,
  inner sep=0.4cm,
  outer sep=0.4cm,
  rounded corners=5pt
]

\newcommand{\pictgraphqqueryacyclic}{
  \begin{tikzpicture}[
      xscale=1.0,
      yscale=1.0,
      ]
            
      \node (X) at (-1,8){$X$};
      \node (U) at (-1,5){$U$};
      \node (U1) at (-4,4){$U_1$};
      \node (U2) at (0.5,4){$U_2$};
      \node (W) at (-4,3){$W$};

      \node (W1) at (-5.5,2){$W_1$};
      \node (W2) at (-2.5,2){$W_2$};
      
      \node (Y1) at (-5.5,1){$Y_1$};
      \node (Y2) at (-2.5,1){$Y_2$};
      \node (Y3) at (1.5,1){$Y_3$};

      \node (x1) at (-9,0)[label={below:$x_1$}]{};
      \node (u1) at (-8,0)[label={below:$u_1$}]{};
      \node (w1) at (-7,0)[label={below:$w_1$}]{};
      \node (y1) at (-6,0)[label={below:$y_1$}]{};
      \node (x2) at (-5,0)[label={below:$x_2$}]{};
      \node (w2) at (-4,0)[label={below:$w_2$}]{};
      \node (y2) at (-3,0)[label={below:$y_2$}]{};
      \node (x3) at (-2,0)[label={below:$x_3$}]{};
      \node (w3) at (-1,0)[label={below:$w_3$}]{};
      \node (u2) at (0,0)[label={below:$u_2$}]{};
      \node (y3) at (1,0)[label={below:$y_3$}]{};
      \node (x4) at (2,0)[label={below:$x_4$}]{};
      \node (u3) at (3,0)[label={below:$u_3$}]{};
      \node (y4) at (4,0)[label={below:$y_4$}]{};

      \draw [uEdge, shorten >=0pt] (X) to (x1.center);
      \draw [uEdge, shorten >=0pt] (X) to (y4.center);
      \draw [snakeEdgea] (X) to (U);
      
      \draw [dEdge] (U) to (U1);
      \draw [dEdge] (U) to (U2);
      
      \draw [uEdge, shorten >=0pt] (U1) to (u1.center);
      \draw [uEdge, shorten >=0pt] (U1) to (u2.center);
      \draw [snakeEdgea] (U1) to (W);
      
      \draw [uEdge, shorten >=0pt] (U2) to (u2.center);
      \draw [uEdge, shorten >=0pt] (U2) to (u3.center);
      \draw [snakeEdgea] (U2) to (Y3);

      \draw [dEdge] (W) to (W1);
      \draw [dEdge] (W) to (W2);      

      \draw [uEdge, shorten >=0pt] (W1) to (w1.center);
      \draw [uEdge, shorten >=0pt] (W1) to (w2.center);
      \draw [snakeEdgea] (W1) to (Y1);

      \draw [uEdge, shorten >=0pt] (W2) to (w2.center);
      \draw [uEdge, shorten >=0pt] (W2) to (w3.center);
      \draw [snakeEdgea] (W2) to (Y2);
      
      \draw [uEdge, shorten >=0pt] (U1) to (u1.center);
      \draw [uEdge, shorten >=0pt] (U1) to (u2.center);
      
      \draw [uEdge, shorten >=0pt] (Y1) to (y1.center);
      \draw [uEdge, shorten >=0pt] (Y1) to (x2.center);
      
      \draw [uEdge, shorten >=0pt] (Y2) to (y2.center);
      \draw [uEdge, shorten >=0pt] (Y2) to (x3.center);
      
      \draw [uEdge, shorten >=0pt] (Y3) to (y3.center);
      \draw [uEdge, shorten >=0pt] (Y3) to (x4.center);
      
      \draw [uEdge, shorten >=0pt, shorten <=0pt] (x1.center) to (y1.center);
      \draw [uEdge, shorten >=0pt, shorten <=0pt] (x2.center) to (y2.center);
      \draw [uEdge, shorten >=0pt, shorten <=0pt] (x3.center) to (y3.center);
      \draw [uEdge, shorten >=0pt, shorten <=0pt] (x4.center) to (y4.center);

      \draw [decorate,decoration={brace,amplitude=10pt, mirror},xshift=-4pt,yshift=0pt] ($(x1)+(-0.2,-0.5)$) to ($(y1)+(0.2,-0.5)$);
      \node (s1) at ($(x1)!0.5!(y1) + (0, -1.0)$){$s_1$};
      
      \draw [decorate,decoration={brace,amplitude=10pt, mirror},xshift=-4pt,yshift=0pt] ($(x2)+(-0.2,-0.5)$) to ($(y2)+(0.2,-0.5)$);
      \node (s2) at ($(x2)!0.5!(y2) + (0, -1.0)$){$s_2$};

      \draw [decorate,decoration={brace,amplitude=10pt, mirror},xshift=-4pt,yshift=0pt] ($(x3)+(-0.2,-0.5)$) to ($(y3)+(0.2,-0.5)$);
      \node (s3) at ($(x3)!0.5!(y3) + (0, -1.0)$){$s_3$};

      \draw [decorate,decoration={brace,amplitude=10pt, mirror},xshift=-4pt,yshift=0pt] ($(x4)+(-0.2,-0.5)$) to ($(y4)+(0.2,-0.5)$);
      \node (s4) at ($(x4)!0.5!(y4) + (0, -1.0)$){$s_4$};

    \end{tikzpicture}%
}

\newcommand{\pictgraphqqueryacyclicb}{
  \begin{tikzpicture}[
      xscale=1.2,
      yscale=0.8,
      ]
            
      \node[mnode] (1) at (0,3){};
      \node[mnode] (1b) at (5,3.3)[label={below:$x_1$}]{};
      \node[mnode] (2) at (1,2)[label={below:$x_2$}]{};
      \node[mnode] (3) at (2,1)[label={below:$z$}]{};
      \node[mnode] (4) at (3,2)[label={below:$z'$}]{};
      \node[mnode] (5) at (4,0)[label={below:$u$}]{};
      \node[mnode] (6) at (5,0)[label={below:$v$}]{};
      \node[mnode] (7) at (6,2)[label={below right:$y_2$}]{};
      \node[mnode] (9) at (8,3)[label={below:$y_1$}]{};
      \node[invisiblel] (1a) at (0,3){$ $};
      \node[invisiblel] (2a) at (1,2){$ $};
      \node[invisiblel] (3a) at (2,1){$ $};
      \node[invisiblel] (4a) at (3,2){$ $};
      \node[invisiblel] (5a) at (4,0){$ $};
      \node[invisiblel] (6a) at (5,0){$ $};
      \node[invisiblel] (7a) at (6,2){$ $};
      \node[invisiblel] (9a) at (8,3){$ $};

      \draw [snakeEdge] (1a) to (1b);
      \draw [snakeEdge] (1b) to (9a);
      \draw [snakeEdge] (1a) to (2a);
      \draw [snakeEdge] (2a) to (3a);
      \draw [dEdge] (3) to (4);
      \draw [snakeEdge] (3a) to (5a);
      \draw [snakeEdge] (4a) to (7a);
      \draw [snakeEdge] (7a) to (9a);
      \draw [snakeEdge] (6a) to (7a);
      \draw [dEdge] (5) to node[below] {$\sigma$} (6) ;;
  \end{tikzpicture}
}

\newcommand{\pictlde}{
  \begin{tikzpicture}[
      xscale=1.2,
      yscale=0.8,
      ]
      \node (x) at (0,0) {$\times$};    
      \node[mnode] (s1) at (-5,0) {};
      \node[mnode] (t1) at (-1,0) {}
      	 edge[->,dashed,bend right=90] (s1) {};
      \node[mnode] (s1) at (-5,0) {}
     	 edge [snakeEdge]  node[midway,below] {$a_1 - 1$} (t1)
     	 edge [snakeEdge, bend right=90]  node[midway,below] {$a_n - 1$} (t1);
      \node (s1b) at (-5.5,0) {$s_1$};
      \node (dots) at (-3,-.75) {$\vdots$};

      \node (s1b) at (-.5,0) {$t_1$};

     \node[mnode] (s2) at (1,0) {};
     \node (s2b) at (1,-.35) {$s_2$};
     \node[mnode] (t2) at (5,0) {};
     \node (t2b) at (5,-.35) {$t_2$};
     \node (s2) at (s2)  {}
     	edge[snakeEdge] node[midway,below] {$b$} (t2);
     
	\node (G1) at (-3,2) {$G_1$};	
	\node (G2) at (3,1) {$G_2$};	
  \end{tikzpicture}
}

\newcommand{\pictreductiontree}{
  \begin{tikzpicture}[
      xscale=1.5,
      yscale=1.0,
    font=\scriptsize, 
      ]
            
      \node[mnode] (00) at (-0.5,2)[label={left:$(0,0)$}]{};
      \node[mnode] (11) at (1,1)[label={below:$(1,1)$}]{};
      \node[mnode] (12) at (1,2)[label={below:$(1,2)$}]{};
      \node[mnode] (1n1) at (1,4)[label={above:$(1,n^2-1)$}]{};
      \node[mnode] (1n) at (1,5)[label={above:$(1,n^2)$}]{};

      \node[mnode] (22) at (2,2)[label={below:$(2,2)$}]{};
      \node[mnode] (2n1) at (2,4)[label={above:$(2,n^2-1)$}]{};
      \node[mnode] (2n) at (2,5)[label={above:$(2,n^2)$}]{};

      \node (3n1) at (3,4){};
      \node (3n) at (3,5){};

      \node (4n1) at (4,4){};
      \node (4n) at (4,5){};

      \node[mnode] (5n1) at (5,4)[label={above:$(n^2-1,n^2-1)$}]{};
      \node[mnode] (5n) at (5,5)[label={above:$(n^2-1,n^2)$}]{};

      \node[mnode] (6n) at (6,5)[label={above:$(n^2,n^2)$}]{};

      \draw [dEdge] (00) to (11);
      \draw [dEdge] (00) to (12);
      \draw [dEdge] (00) to (1n1);
      \draw [dEdge] (00) to (1n);
      
      \draw [dEdge] (12) to (22);
      \draw [dEdge] (1n1) to (2n1);
      \draw [dEdge] (1n) to (2n);
      
      \draw [dEdge] (2n1) to (3n1);
      \draw [dEdge] (2n) to (3n);
      
      \draw [dotted, thick] (3n1) to (4n1);
      \draw [dotted, thick] (3n) to (4n);
      
      \draw [dEdge] (4n1) to (5n1);
      \draw [dEdge] (4n) to (5n);
      
      \draw [dEdge] (5n) to (6n);

      \draw [dotted, thick] (12) to (1n1);
      \draw [dotted, thick] (22) to (2n1);

  \end{tikzpicture}

} %

   \newcommand{\shortVersion}[1]{}
   \newcommand{\longVersion}[1]{#1}

  \maketitle
  \begin{abstract}

Graph databases in many applications---semantic web, transport or biological networks among others---are not only large, but also frequently modified. Evaluating graph queries in this dynamic context is a challenging task, as those queries often combine first-order and navigational features. 

Motivated by recent results on maintaining dynamic reachability, we study the dynamic evaluation of traditional query languages for graphs in the descriptive complexity framework.
Our focus is on maintaining regular path queries, and extensions thereof, by first-order formulas. In particular we are interested in path queries defined by non-regular languages and in extended conjunctive regular path queries (which allow to compare labels of paths based on word relations). Further we study the closely related problems of maintaining distances in graphs and reachability in product graphs.

In this preliminary study we obtain upper bounds for those problems in restricted settings, such as undirected and acyclic graphs, or under insertions only, and negative results regarding quantifier-free update formulas. In addition we point out interesting directions for further research.

  \end{abstract}

  \section{Introduction}\label{section:introduction}
\makeatletter{}%
Graph databases are important in applications in which the topology of data is as important as the data itself. 
Intuitively, a graph database represents objects (by nodes), and relationships between those objects (often modeled by labeled edges---see \cite{angles2008survey} for a survey on graph database models). The last years have witnessed an increasing interest in graph databases, due to the uprise of applications that need to manage and query massive and highly-connected data, as for example the semantic web, social networks or biological networks. In most of these applications, databases are not only large, but also highly dynamic. Data is frequently inserted and deleted, and hence so is its network structure. The goal of this work is to explore how query languages for graph databases can be evaluated in this dynamic context.

Many query languages for graph databases combine traditional first-order features with {\em navigational} ones. Already basic languages (such as regular path queries, see e.g. \cite{Wood12survey, Baeza13}) allow to test the existence of paths satisfying constraints on their labels (e.g. adherence to a regular expression in regular path queries). Computing  the answers to this kind of queries on large, highly dynamic graphs is a big challenge. It is conceivable, though, for answers to a query before and after small modifications to be closely related. Thus a reasonable hope is to be able to update the answer to a query in a more efficient way than recomputing it from scratch after each modification. Even more so if we allow to store extra auxiliary data that might ease the updating task. To what extent this is possible, and in which precise conditions, is the subject of {\em dynamic computational complexity}.

Here we are interested in studying the dynamic complexity of query languages for graph databases from a 
 {\em descriptive} approach. In the dynamic descriptive complexity setting, proposed independently by Dong, Su and Topor \cite{DongT92, DongS93} and by Patnaik and Immerman \cite{PatnaikI97}, a {\em dynamic program} maintains auxiliary relations with the intention to help answering a query  over a (relational) database subject to small modifications (insertions or deletions of tuples). When a modification occurs, the query answers and every auxiliary relation are updated by first-order formulas (or, equivalently, by core SQL queries) evaluated over the current database and the available auxiliary data. Such programs benefit therefore from being both highly parallelizable (due to the close connection of first-order logic and small depth boolean circuits) and readily implementable in standard relational database engines. The class of queries maintainable by first-order update formulas is called \DynFO. 

Query languages for graphs have, so far, not been studied systematically in the dynamic descriptive complexity setting. Very likely the main reason is that until recently it was not even known whether reachability in directed graphs could be maintained by first-order update formulas. That this indeed is possible was shown in \cite{DattaKMSZ15}, with the immediate consequence that all fixed (conjunctions of) regular path queries can also be maintained. Thus regular path queries can be evaluated in a highly parallel fashion in dynamic graph databases.%

Motivated by this result we study the dynamic maintainability of more expressive query languages.

\begin{goal*}
  Gain a better understanding of the limits of maintaining graph query languages in the dynamic context.
\end{goal*}

Our focus is on regular path queries and extensions thereof---non-regular path queries and extended conjunctive regular path queries (short: ECRPQs).

Some previous work on non-regular path queries has been done. Weber and Schwentick exhibited a context-free path query (the Dyck language $D_2$) that can be maintained in $\DynFO$ on acyclic graphs \cite{WeberS07}. Also, for the simple class of path-shaped graph databases, formal language results can be transferred. Already Patnaik and Immerman pointed out that regular languages can be maintained in $\DynFO$~\cite{PatnaikI97}. Later, Gelade et al.\ systematically studied the dynamic complexity of formal languages \cite{GeladeMS12}. They showed, among other results, that regular languages can be maintained by quantifier-free update formulas, and that all context-free languages can be maintained in $\DynFO$. 

The second extension of regular path queries to be studied here are extended conjunctive regular path queries. In previous work it has been noticed that conjunctions of regular path queries (CRPQs) fall short in expressive power for modern applications of graph databases~\cite{BarceloLLW12}. A feature commonly demanded by these applications is the comparison of labels of paths defined by CRPQs based on relations of words (e.g. prefix, length constraints, fixed edit-distance). ECRPQs have been introduced to fulfill this requirement \cite{BarceloLLW12}, that is, they generalize CRPQs by allowing to test whether multiple labels of paths adhere to given regular relations. Two basic properties expressible by ECRPQs are whether two pairs of nodes are connected by paths of the same length and if so, whether also paths with the same label sequence exist. In general, maintaining the result of ECRPQs seems to be a difficult task. In this article we therefore explore the maintenance of ECRPQs in restricted settings. 

Finally, there is also a close connection between the evaluation of graph queries and the reachability problem in unlabeled and labeled product graphs. We discuss this connection (see Section \ref{section:preliminaries}), and exploit it in several of our results.

\subparagraph*{Contributions}

First we study path queries and show that
\begin{itemize}
 \item all regular path queries can be maintained by quantifier-free formulas when only insertions are allowed,
 \item all context-free path queries can be maintained by first-order formulas on acyclic graphs, and 
 \item there are non-context-free path queries maintainable by first-order formulas on undirected and acyclic graphs, as well as on general graphs under insertions only.
\end{itemize}

As a first step towards maintaining 
ECRPQs we explore for which graph classes the lengths of paths between nodes can be maintained.  We exhibit dynamic programs for maintaining all distances for undirected and acyclic graphs, as well as for directed graphs when only insertions are allowed. It remains open, whether distances can be maintained in $\DynFO$ for general directed graphs, but we show that quantifier-free update formulas do not suffice. 

The techniques used to maintain all distances can be used to maintain variants of ECRPQs in restricted settings. Denote the extension of a class of queries by linear constraints on the number of occurrences of symbols on paths by +LC. This extension was introduced and studied in \cite{BarceloLLW12}. We show that
\begin{itemize}
 \item all CRPQ+LCs can be maintained by first-order formulas when only insertions are allowed, and
 \item all ECRPQ+LCs can be maintained by first-order formulas on acyclic graphs. 
\end{itemize}

An immediate consequence of our results for distances is that reachability can be maintained in products of (unlabeled) graphs for those restrictions. By using the dynamic program for maintaining the rank of matrices from \cite{DattaKMSZ15}, we extend this result to more general graph products. Furthermore we show that pairs of nodes connected by paths with the same label sequence can be maintained in acyclic graphs using first-order update formulas. 

\subparagraph*{Related work} The maintenance of problems has also been studied from an algorithmic point of view. A good starting point for readers interested in upper bounds for dynamic algorithms is \cite{RodittyZ08, DemetrescuI08}; a good starting point for lower bound techniques is  the survey by Miltersen on cell probe complexity~\cite{Miltersen99}. The upper bounds for reachability obtained in  \cite{RodittyZ08, DemetrescuI08} immediately transfer to dynamic algorithmic evaluation of regular path queries (using the reduction exhibited in~\cite{DattaKMSZ15}).

\subparagraph*{Outline} The dynamic setting and the basic graph query languages are introduced in Section~\ref{section:preliminaries}. There we also discuss the connection between query evaluation and reachability in product graphs. Section \ref{section:pathqueries} contains the results on maintaining graph queries. Our results for maintaining distances and ECRPQs are presented in Section \ref{section:beyondpathqueries}. In Section~\ref{section:reachabilityInProductGraphs} some of the results for maintaining graph queries are transferred to reachability in graph products, and we also provide results for reachability in generalized graph products. We conclude in Section \ref{section:conclusion}.

This is a full version of \cite{MunozVZ16}.

\subparagraph*{Acknowledgements}
We thank Pablo Barcel\'o, Samir Datta and Thomas Schwentick for stimulating and illuminating discussions.

%
%
%
%
%
%
%
%
%
%
%
%
%
%
%
%
%
%
%
%
%
%
%
%
%
%
%
%
%
%
%
%
%
%
%
%
%
%
%
%
%
%
%
%
%
%
%
%

%
%
%
%
%
%
%
%
%
%
%
%
%
%

%
%
%

%

%
%

%
%
%
%
%
%
%
%
%
%
%
%
%
%
%
%
%
%
%
%
%
%
%
%
%
%
%
%
%
%
%
%
%
%
%
%
%
%
%
%
%
%
%
%
%
%
%
%
%
%
%
%
%
%
%
%
%
%
%
%
%
%
%
%
%
%
%
%
%
%
%
%
%
%
%
%
%
%
%
%
%
%
%
%
%
%
%
%
%
%
%
%
%
%
%
%
%
%
%
%
%
%
%
%
%
%
%
%
%
%
%
%
%
%
%
%
%
%
%
%
%
%
%
%
%
%
%
%
%
%
%
%
%
%
%

  \section{Preliminaries}\label{section:preliminaries}
   \toAppendix{\section{Proofs and Details for Section \ref{section:preliminaries}}}
\makeatletter{}%
In this section we introduce the dynamic complexity framework as well as the graph query languages used in this article.

\subparagraph*{Dynamic complexity framework}

In this work we use the dynamic complexity framework as introduced by Patnaik and Immerman \cite{PatnaikI97}. 
The following introduction of the framework is borrowed from previous work \cite{ZeumeS15reach}.

Intuitively, the goal of a dynamic program is to keep the result of a given query $\query$ up to date while the database to be queried (the \emph{input database}) is subject to tuple insertions and deletions. To this end the dynamic program stores auxiliary relations (the \emph{auxiliary database}) with the aim that one of those relations always (that is, after every possible sequence of modifications), stores the result of $\query$ for the current input structure. Whenever a tuple is inserted into or deleted from the input structure, each auxiliary relation is updated by the dynamic program by evaluating a specified first-order formula.

We make this more precise now. A \emph{dynamic instance}\defindex{dynamic instance} of a query $\query$ is a pair $(\db, \alpha)$, where $\db$ is a database over some finite domain $\domain$ and $\alpha$ is a sequence of modifications to~$\db$. Here, a \emph{modification}\defindex{modification} is either an insertion of a tuple over $\domain$ into a relation of~$\db$ or a deletion of a tuple from a relation of~$\db$. The result of $\query$ for $(\db, \alpha)$ is the relation that is obtained by first applying the modifications \mbox{from $\alpha$} to $\db$ and then evaluating $\query$ on the resulting  database. We use the Greek letters $\alpha$ and $\beta$ to denote modifications as well as modification sequences.
The database resulting from applying a modification $\alpha$ to a database $\db$ is denoted by $\alpha(\db)$. The result $\updateDB{\alpha}{\db}$ of applying a sequence of modifications $\alpha \df \alpha_1 \ldots \alpha_m$ to a database $\db$ is defined by~$\updateDB{\alpha}{\db} \df \updateDB{\alpha_m}{\ldots (\updateDB{\alpha_1}{\db})\ldots}$.

Dynamic programs, to be defined next, consist of an initialization mechanism and an update program.  The former  yields, for every (input) \mbox{database $\db$},  an initial state with initial auxiliary  data. The latter defines how the new state of the dynamic program is obtained from the current state when applying a modification.

A \emph{dynamic schema}\defindex{dynamic schema} is a tuple $(\inpSchema, \auxSchema)$ where $\inpSchema$\defindex{$\inpSchema$}\defindex{input schema} and $\auxSchema$\defindex{$\auxSchema$}\defindex{auxiliary schema} are the schemas of the input database and the auxiliary database, respectively. While $\inpSchema$ may contain constants, we do not allow constants in $\auxSchema$ in the basic setting. We always let $\tau\df\inpSchema\cup\auxSchema$. 

\defindex{update program}\defindex{update formula}
\begin{definition}[Update program]\label{def:updateprog}
  An \emph{update program} $P$ over a dynamic schema \mbox{$(\inpSchema, \auxSchema)$} 
  is a set of first-order formulas (called \textit{update formulas} in the following) that contains,  for every relation symbol $R$ in $\auxSchema$ and every $\delta \in \{\ins_S, \del_S\}$ with $S \in \inpSchema$, an update formula  $\uf{R}{\delta}{\bar x}{\bar y}$\defindex{$\uf{R}{\delta}{\bar u}{\bar x}$} over the schema~$\schema$  where $\bar x$ and $\bar y$ have the same arity as $S$ and $R$, respectively.
\end{definition}

A \emph{program state}\defindex{program state}\index{state|see {program state}} $\state$ over dynamic schema \mbox{$(\inpSchema, \auxSchema)$} is a structure $(\domain, \inp,  \aux)$ where $\domain$ is a finite domain, $\inp$ is a database over the input schema (the \emph{current database}) and $\aux$ is a database over the auxiliary schema (the \emph{auxiliary database}\defindex{auxiliary database}).

The semantics of update programs is as follows\defindex{absolute semantics}. Let $P$ be an update program, $\state=(\domain, \inp,\aux)$ be a program state and $\alpha = \delta(\bar a)$ a modification where $\bar a$ is a tuple over $\domain$ and \mbox{$\delta \in \{\ins_S, \del_S\}$} for some $S \in \inpSchema$. If $P$ is in state $\state$ then the application of $\alpha$ yields the new state $\updateState{\prog}{\alpha}{\state} \df (\domain, \alpha(\inp), \aux')$  where, in $\aux'$, a relation symbol $R \in \auxSchema$ is interpreted by $\{\bar b \mid \state \models \uf{R}{\delta}{\bar a}{\bar b}\}$. The effect $P_\alpha(\state)$\defindex{P@$P_\alpha(\state)$} of applying a modification sequence $\alpha \df \alpha_1 \ldots \alpha_m$ to a state $\state$ is the state $\updateState{P}{\alpha_m}{\ldots (\updateState{P}{\alpha_1}{\state})\ldots}$. 

\defindex{dynamic program}
\begin{definition}[Dynamic program] \label{definition:dynprog}
  A \emph{dynamic program} is a triple $(P,\init,Q)$, where
  \begin{compactitem}
   \item  $P$ is an update program over some dynamic schema
  \mbox{$(\inpSchema, \auxSchema)$}, 
    \item \init is a mapping that maps $\inpSchema$-databases to $\auxSchema$-databases, and 
    \item $Q\in\auxSchema$ is a designated \emph{query symbol}\defindex{query symbol}.
  \end{compactitem}
\end{definition}

A dynamic program $\calP=(P,\init,Q)$ \emph{maintains} \index{maintenance of a query}  a query  $\query$ if, for every dynamic instance $(\db,\alpha)$, the query result $\query(\alpha(\db))$ coincides with the content of $Q$ in the state \mbox{$\state=P_\alpha(\state_\init(\db))$} where $\state_\init(\db)$ is the initial state for $\db$, that is, \mbox{$\state_\init(\db) \df (\domain, \db,  \init(\db))$}.

The following example due to \cite{PatnaikI97} shows how the transitive closure of an acyclic graph subject to edge insertions and deletions can be maintained in this set-up. The basic technique of this example will be crucial in some of the later proofs. 

\begin{example}\label{example:acyclictc}
  Consider an acyclic graph $G$ subject to edge insertions and deletions. In the following, our goal is to maintain the transitive closure of $G$ using a dynamic program with first-order update formulas. %
  It turns out that if the graph is guaranteed to remain acyclic, then it is sufficient to store the current transitive closure relation in an auxiliary relation $T$. We follow the argument from \cite{PatnaikI97}.
   
   When an edge $(u, v)$ is inserted into $G$ the following very simple rule updates $T$: there is a path from $x$ to $y$ after inserting $(u, v)$ if (1) there was already a path from $x$ to $y$ before the insertion, or (2) there were paths from $x$ to $u$ and from $v$ to $y$ before the insertion. This rule can be easily specified by a first-order update formula that defines the updated transitive closure relation\footnote{For simplicity we use the same names for elements and variables.}: $\uf{T}{\ins_{E}}{u,v}{x,y} \df T(x,y) \vee \big(T(x,u) \wedge T(v,y)\big)$. 

   Deletions are slightly more involved. There is a path~$\rho$ from $x$ to $y$ after deleting an edge $(u,v)$ if there was a path from $x$ to $y$ before the deletion and (1) there was no such path via $(u,v)$, or (2) there is an edge $(z,z')$ on $\rho$ such that $u$ can be reached from $z$ but not \mbox{from $z'$}. If there is still a path $\rho$ from $x$ to $y$, such an edge $(z,z')$ must exist, as otherwise~$u$ would be reachable from $y$, contradicting acyclicity. This rule can be described by a first-order formula:
  \begin{multline*}
       \uf{T}{\del_{E}}{u,v}{x,y} \df T(x,y) \wedge \Big( \big(\neg T(x,u) \vee \neg T(v,y)\big) \vee \exists z \exists z'\\
      \quad \quad \big(T(x,z) \wedge E(z,z') \wedge (z \neq u \vee z' \neq v) \wedge T(z',y) \wedge T(z,u) \wedge \neg T(z',u)\big)\Big)
  \end{multline*}%
 \qed%
\end{example}
A word on the initial input and auxiliary databases is due. As default we use the original setting of Patnaik and Immerman, where the input database is empty at the beginning, and the auxiliary relations are initialized by first-order formulas evaluated on the initial input database. When we use a different initialization setting we state it explicitly. In the literature several other settings have been investigated and we refer to \cite{ZeumeS15reach, Zeume15thesis} for a detailed discussion.  

The class of queries that can be maintained by first-order update formulas in the setting of Patnaik and Immerman is called\footnote{In \cite{ZeumeS15reach, ZeumeS14cq, Zeume14clique} the class $\DynFO$ comes with an arbitrary initialization, yet there the focus is on lower bounds.} $\DynFO$. Restricting update formulas to be quantifier-free yields the class $\DynProp$.

When showing that a particular query is in $\DynFO$ we often assume that \emph{arithmetic} on the domain is available from initialization time, that is, we assume the presence of relations $\leq, +, \times$ that are interpreted as a linear order---allowing to identify elements with numbers---, addition and multiplication on the domain. From a $\DynFO$ program that relies on built-in arithmetic, a program without built-in arithmetic can be constructed for all queries studied here by using a technique from \cite{DattaKMSZ15}.

\begin{proposition}[{\cite[Theorem 4]{DattaKMSZ15}}]\label{prop:removing-arithmetic}
Every domain-independent query $\query$ that can be maintained in \DynFO with built-in arithmetic can also be maintained in \DynFO.
\end{proposition}
Here, a query is \emph{domain-independent} if its result does not change when elements are added to the domain.%

Constructing a $\DynFO$ program for a specific query $\query$ can be a tedious task. Such a construction can often be simplified by reducing $\query$ to a query $\query'$ for which a dynamic program has already been obtained. Such a reduction needs to be consistent with first-order logic and its use in this dynamic context. A suitable kind of reductions are \emph{bounded first-order reductions}. Intuitively, a query $\query$ reduces to a query $\query'$ via a bounded first-order reduction if a modification of an instance of $\query$ induces constantly many, first-order definable modifications in a instance of $\query'$. Note that if $\query$ can be reduced to $\query'$ via a bounded first-order reduction, then first-order update formulas for a modification of an instance for $\query$ can be obtained by composing the first-order update formulas for the corresponding (first-order definable) modications of the instance of $\query'$. We refer to \cite{PatnaikI97} and  \cite{GraedelS12} for a detailed exposition to bounded first-order reductions.

In this article we study dynamic programs for queries on (labeled) graphs. For most of our dynamic programs the precise encoding of graphs is not important. If the input to a query is a single $\Sigma$-labeled graph $G = (V,E)$   then it can, for example, be encoded by binary relations $E_\sigma$ that store all $\sigma$-labeled edges, for all $\sigma \in \Sigma$. Similarly for constantly many graphs. Some of our results are for input databases that contain more than constantly many graphs. Those can be encoded in higher arity relations in a straightforward way. For example, linearly many graphs can be stored in ternary relations $E_\sigma$ containing a tuple $(g, u, v)$ if graph $g$ contains a $\sigma$-labeled edge $(u,v)$.

\subparagraph*{Graph databases and query languages}
We review basic definitions of graph databases in order to fix notations and introduce the query languages used in this work.

A \emph{graph database} over an alphabet $\Sigma$ is a finite $\Sigma$-labeled graph $G = (V, E)$ where $V$ is a finite set of nodes and $E$ is a set of labeled edges $(u, \sigma, v) \subseteq V \times \Sigma \times V$. Here $\sigma$ is called the \emph{label} of edge $(u, \sigma, v)$. %
Given a $\Sigma$-labeled graph $G=(V,E)$ and a symbol $\sigma \in \Sigma$, we denote by $G_\sigma$ the projection of $G$ onto its $\sigma$-labeled edges, that is, the graph $G_\sigma$ has the edge set $\{(u,v) \mid (u,\sigma,v) \in E\}$. We say that a $\Sigma$-labeled graph $G$ is {\em acyclic} if the graph $\cup_{\sigma \in \Sigma} G_\sigma$ is acyclic, and {\em undirected}, if for each $\sigma \in \Sigma$ the graph $G_\sigma$ is undirected. 

A {\em path} $\rho$ in $G$ from $v_0$ to $v_m$ is a sequence of edges $(v_0,\sigma_1,v_1),\ldots,(v_{m-1},\sigma_m,v_m)$ of $G$, for some length $m \geq 0$. 	The {\em label}
of $\rho$, denoted by $\lambda(\rho)$, is the word $\sigma_1 \cdots \sigma_m \in \Sigma^*$. Paths of length zero are labeled by the empty string $\epsilon$. For a formal language $L \subseteq \Sigma^*$, we say that $\rho$ is an $L$-path if $\lambda(\rho) \in L$.

The basic building block of many graph query languages are {\em regular path queries} (short: RPQs). An RPQ selects all pairs of nodes in a $\Sigma$-labeled graph that are connected by an $L$-path, for a a given regular language $L \subseteq \Sigma^*$. Here we are interested in two extensions of regular path queries. One of them are path queries defined by non-regular languages, namely context-free and non-context-free languages.

The second extension to be studied, \emph{extended  conjunctive regular path queries} (short: ECRPQs), allows to define multiple paths and  to compare their labels based on relations on words.
In the following we give a short introduction to ECRPQs and refer to \cite{BarceloLLW12} for a detailed study. 

In ECRPQs, paths are compared by {\em regular relations}. A $k$-ary regular relation $R$ over alphabet $\Sigma$ is defined by a finite state automaton $\calA$ that synchronously reads $k$ words over $\Sigma \cup \bot$, with $\bot \notin \Sigma$ . The $\bot$ symbol is a padding symbol that may only occur at the end of a word, and therefore allows for processing words of different length.  More formally $\calA$ reads words over the alphabet $(\Sigma\cup{\bot})^k$, and a $k$-tuple of words is in $R$ if its corresponding string over  $(\Sigma\cup{\bot})^k$ is accepted by $\calA$. 

An ECRPQ is of the form 
$\query(\vec z) \longleftarrow \bigwedge_{1 \leq i \leq m} (x_i, \pi_i, y_i), \bigwedge_{1 \leq j \leq t} R_j(\vec \omega _j)$
where
\begin{itemize}
\item each $R_j$ is a regular relation over $\Sigma$ (specified by some finite state automaton),
\item $\vec x = (x_1,...,x_m)$, $\vec y = (y_1,...,y_m)$ and $\vec z$ are tuples of node variables such that the variables in $\vec z$ occur in $\vec x$ or $\vec y$, and
\item $\vec \pi = (\pi_1,...,\pi_m)$ and $\vec \omega_1,...,\vec \omega_t $ are distinct tuples of path variables such that all variables in each $\vec \omega_j$ occur in $\vec \pi$.
\end{itemize}

In general, both node and path variables can occur in the head of an ECRPQ. \shortOrLong{Here we focus on ECRPQs with heads containing node variables only.}{Outputs of ECRPQs are potentially infinite sets then, since there can be infinitely many paths in graphs with cycles. Nevertheless, given a $\Sigma$-labeled graph $G = (V,E)$ and a tuple $\vec v$ of nodes, the answer set is a regular relation over the alphabet $(V^k \cup \Sigma_\bot^k)$,
where $k$ is the  number of path variables in the head. 
Such a regular relation is used as an encoding of all possible path outputs.
For a fixed ECRPQ $\query$, given $G$ and $\vec v$, the automaton for this regular relation can obtained in $\PTIME$ \cite{BarceloLLW12}. More precisely, it
is definable by first-order queries evaluated on the graph database.  We can therefore neglect path variables
in the dynamic setting---for every tuple of nodes in the answer of the query we can obtain the regular relation
encoding the output paths by first order queries.}

The semantics of ECRPQs is defined in a natural way. For an ECRPQ $\query$ of the above form, a $\Sigma$-labeled graph $G = (V,E)$, and mappings $\nu$ from node variables to nodes and $\mu$ from path variables to paths, we write $(G,\nu, \mu) \models \query$ if
\begin{itemize}
\item $\mu(\pi_i)$ is a path in $G$ from $\nu(x_i)$ to $\nu(y_i)$ for $ 1 \leq i \leq m$, and
\item the tuple $(\lambda(\mu(\pi_{j_1})), ..., \lambda(\mu(\pi_{j_k})) )$ belongs to the relation $R_j$ for each $\vec \omega_j = (\pi_{j_1},...,\pi_{j_k})$.
\end{itemize}
The result of $\query$ evaluated on $G$ is defined by $Q(G) \df \{\nu(\vec z) : (G,\nu, \mu) \models \query\}$. 

\subparagraph*{Product Graphs and Graph Query Languages} There is a strong connection between the evaluation problem for many graph query languages and the reachability query for products of labeled graphs. For example, the evaluation of a regular path query $L$ on a labeled graph $G$ can be reduced to reachability in the product graph $\calA \times G$ where $\calA$ is a finite state automaton for $L$. Product graphs also help for the evaluation of fragments of ECRPQs as well. We will exploit this connection at several places and therefore present some basic properties of product graphs next.

The {\em product graph} $\prod_i G_i$ of $m$ $\Sigma$-labeled graphs $G_i = (V_i,E_i)$,  $1 \leq i \leq m$,  has nodes $\prod_i V_i$ and an edge $(\vec x, \vec y)$ between two nodes $\vec x = (x_1,...,x_m)$ and $\vec y  = (y_1,...,y_m)$ if there is a symbol $\sigma \in \Sigma$ such that $(x_i, \sigma, y_i) \in E_i$ for each $1 \leq i \leq m$. The graphs $G_i$ are called \emph{factors} of the graph product. Graph products for unlabeled graphs are defined analogously. The following well known property characterizes reachability in (labeled) product graphs.

\begin{fact}\label{prop:reach-prod-dists}
Let $(G_i)_{1 \leq i \leq m}$ be graphs ($\Sigma$-labeled graphs) with $G_i = (V_i, E_i)$  and let $\vec x = (x_1, \ldots, x_m), \vec y = (y_1, \ldots, y_m)$ be two pairs of nodes of $\prod_i G_i$. Then $\vec y$ is  reachable from $\vec x$ in $\prod_i G_i$ if and only if there are paths $\rho_i$ from $x_i$ to $y_i$ in $G_i$ with $|\rho_i| \leq |\prod_i V_i|$, for $i \in \{1, \ldots, m\}$, and $|\rho_i| = |\rho_j|$ ($\lambda(\rho_i) = \lambda(\rho_j)$ respectively) for all $i, j \in \{1, \ldots, m\}$. \qed
\end{fact}

The preceding fact can be used in the dynamic context as well, i.e. it is compatible with bounded first-order reductions. More precisely, reachability in products of unlabeled graphs can be inferred from all distances in the factors.  We say that {\em all distances up to $n^c$}, for $c \in \N$, are computed by a dynamic program if, for a graph $G$ with $n$ nodes and arithmetic on the domain\footnote{We note that from the arithmetic on the domain, arithmetic upto $n^c$ can be defined using first-order formulas.}, it maintains a relation $D$ that contains all tuples $(x,y , \ell)$ such that there is a path from $x$ to $y$ of length $\ell$, for $0 \leq \ell \leq n^c$.

\aproposition{theorem:reductions:graphs}{
  The following problems are equivalent under bounded first-order reductions with built-in arithmetic:
  \begin{enumerate}
    \item Maintaining all distances up to $n^2$.
    \item Maintaining reachability in the product of two graphs (both of them subject to modifications).
    \item Maintaining reachability in the product of two graphs, one of them a fixed path.
  \end{enumerate}
  }

 \toLongAndAppendix{
\begin{proofsketch}
Problem (c) is clearly a special case of Problem (b). The reduction from Problem (b) to Problem (a) is an immediate consequence of Fact \ref{prop:reach-prod-dists}: two nodes $(x, x')$ and $(y, y')$ of a product graph $G \times G'$ are connected if and only if there are equal-length paths of length at most $n^2$ from $x$ to $y$ in $G$ and from $x'$ to $y'$ in $G'$. 
  
  Thus it remains to reduce Problem (a) to Problem (c). For a graph $G$ over domain $\domain \df \{0, \ldots, n-1\}$, consider the product graph $G \times P$  where $P$ is the path $\{(0,1), \ldots, (n^2-1, n^2)\}$ (as usual numbers larger than $n$ are encoded as tuples over $\domain$). Then there is a path of length $\ell$ between two nodes $x$ and $y$ of $G$ if and only if there is a path from $(x, 0)$ to $(y, \ell)$ in $G \times P$. Furthermore, the path $P$ is never modified. 
\end{proofsketch}
}

 A similar equivalence can be established for problems related to reachability in products of $\Sigma$-labeled graphs:
 
\aproposition{theorem:reductions:unlabeledgraphs}{
  The following problems are equivalent under bounded first-order reductions:
  \begin{enumerate}
    \item Maintaining the existence of equally labeled paths between two pairs of nodes.
    \item Maintaining reachability in the product of two $\Sigma$-labeled graphs.
    \item Maintaining reachability in the product of two $\Sigma$-labeled graphs, one of them undirected.
    \item Maintaining the palindrome path query on $\Sigma$-labeled graphs.
  \end{enumerate}
  }

\toLongAndAppendix{
\begin{proofsketch}
The equivalence of problems (a) and (b) is an immediate consequence of Fact~\ref{prop:reach-prod-dists}.

  We next show the equivalence of Problems (b) and (c). Clearly, Problem (c) is a special case of (b). For reducing Problem (b) to Problem (c) consider two (directed) $\Sigma$-labeled graphs $G_1$ and $G_2$. Let $\# \notin \Sigma$ be a fresh symbol, and denote $\Sigma_\# = \Sigma \cup\{\#\}$. In a first step, from $G_1$ and $G_2$ we construct two undirected $\Sigma_\#$-labeled graphs $G'_1$ and $G'_2$ such that (1) there is a path between two nodes of $G_1 \times G_2$ if and only if there is a $(\Sigma\circ\{\#\})^*$-labeled path between the corresponding nodes in $G'_1 \times G'_2$, and (2) one modification in $G_i$ corresponds to at most two modifications in $G'_i$ definable in first-order. To this end, the graph $G'_i$ has two nodes $x_{\text{in}}$ and $x_{\text{out}}$ for each node $x$ of $G_i$. An edge $(x,\sigma, y)$ of $G_i$ is encoded by the edges $(x_{\text{out}}, \sigma, y_{\text{in}})$ and $(y_{\text{in}}, \#, y_{\text{out}})$ in $G'_i$. (In particular, the edge $(y_{\text{in}}, \#, y_{\text{out}})$ is present in $G'_i$ as soon as $y$ has an incoming edge in $G_i$.)

  Now every path in $G_1 \times G_2$ corresponds to a $(\Sigma\circ\{\#\})^*$-labeled path in  $G'_1 \times G'_2$, which in turn corresponds to a path in the product graph $G'_1 \times G'_2 \times \calA$ where $\calA$ is the labeled (directed) graph \scalebox{0.6}{\tikz[baseline={([yshift=-1.0ex]current bounding box.center)}]{      
      \node[mnode] (0) at (0,0){};
      \node[mnode] (1) at (1,0){};
      \draw[dEdge]  (0)  to [dEdge,bend left=20] node[midway,above] {$\Sigma$}(1) ;
      \draw[dEdge]  (1)  to [dEdge,bend left=20] node[midway,below] {$\#$}(0) ;
  }}
  representing the language $(\Sigma\circ\{\#\})^*$. Since $G'_1 \times G'_2 \times \calA$ is the product of the undirected graph $G'_1$ and the directed graph $G'_2 \times \calA$, this yields the intended reduction (as one modification in $G_2$ yields at most six modifications in $G'_2 \times \calA$).
  
  We now show that the problems  (b) and (d) are equivalent. For reducing (d) to (b) we consider, for simplicity, only palindromes of even length; the construction can be easily adapted to arbitrary palindromes. Let $G$ be a labeled (directed) graph. Then there is a path from $x$ to $y$ labeled by a palindrome $ww^R$ if and only if there is a node $z$ such that there are $w$ and $w^R$ labeled paths from $x$ to $z$ and from $z$ to $y$, respectively. Thus finding a palindromic path from $x$ to $y$ corresponds to finding a node $z$ such that there is a path from $(x,y)$ to $(z,z)$ in the product graph $G \times G^-$, where $G^-$ denotes the graph obtained from $G$ by reversing each of its edges (definable in first-order). Note that one modification of $G$ corresponds to two modifications in the factors of $G \times G^-$.
  
  For the other direction, let $G_1$ and $G_2$ be two arbitrary $\Sigma$-labeled graphs and let $\#\not\in\Sigma$ be a fresh symbol. We assume, without loss of generality, that the node sets of the graphs are disjoint. There is a path from $(x_1, x_2)$ to $(y_1, y_2)$ in $G_1 \times G_2$ if and only  if there is a word $w$, a $w$-labeled path from $x_1$ to $y_1$ and a $w$-labeled path from $x_2$ to $y_2$. The latter condition is equivalent to the existence of a $w \# w^R$-labeled path in the graph $G \df G_1 \cup G_2^-$ extended by the edge $(y_1, \#, y_2)$. 
\end{proofsketch}
}
%
%
%

%
%
%
%
%
%
%
%
%
%
%
%
%
%
%
%
%
%
%
%
%
%
%
%
%
%
%
%
%
%
%
%
%
%
%
%
%
%
%
%
%
%
%
%
%
%
%
%
%
%
%
%
%
%
%
%
%
%
%
%
%
%
%
%
%
%
%
%
%
%
%
%
%
%
%
%
%
%
%
%
%
%
%
%
%
%
%
%
%
%
%
%
%
%
%
%
%
%
%
%
%
%
%
%
%
%
%
%
%
%
%
%
%
%
%
%
%
%
%
%
%
%
%
%
%
%
%
%
%
%
%
%
%
%
%
%
%
%
%
%
%
%
%
%
%
%
%
%
%
%
%
%
%
%
%
%
%
%
%
%
%
%
%
%
%
%
%
%
%
%
%
%
%
%
%
%
%
%
%
%
%
%
%
%
%
%
%
%
%
%
%
%
%
%
%
%
%
%
%
%
%
%
%
%
%
%
%
%
%
%
%
%
%
%
%
%
%
%
%
%
%
%
%
%
%
%
%
%
%
%
%
%
%
%
%
%
%
%
%
%
%
%
%
%
%
%
%
%
%
%
%
%
%
%
%
%
%
%
%
%
%
%
%
%
%
%
%
%
%
%
%
%
%
%
%
%
%
%
%
%
%
%
%
%
%
%
%
%
%
%
%
%
%
%
%
%
%
%
%
%
%
%
%
%
%
%
%
%
%
%
%
%
%
%
%
%
%
%
%
%
%
%

%
  
 %
 %

  %
  %

%

  \section{Dynamic Path Queries}\label{section:pathqueries}
  \toAppendix{\section{Proofs for Section \ref{section:pathqueries}}}
\makeatletter{}%
Path queries, as mentioned in the introduction, have almost not been studied in dynamic complexity before. Until recently not even the simple query induced by the language $L(a^*)$ was known to be in $\DynFO$. Yet as an immediate consequence of the dynamic first-order update program for reachability exhibited in \cite{DattaKMSZ15}, all fixed regular path queries (and, since $\DynFO$ is closed under conjunctions, also conjunctions of them) can be maintained by first-order update formulas. 

In this section we continue the exploration of the dynamic maintainability of path queries.  We show that under insertions quantifier-free update formulas are sufficient to maintain (fixed) regular path queries, and that more expressive path queries can be maintained for restricted classes of graphs and constrained modifications.

\atheorem{theorem:graphqueries:regular}{
  When only insertions are allowed then every regular path query can be maintained by quantifier-free update formulas.
}
We conjecture that quantifier-free update formulas do not suffice to maintain RPQs under both insertions and deletions. This would imply that reachability can be maintained without quantifiers which seems to be very unlikely. A first step towards verifying this conjecture was done in \cite{ZeumeS15reach} where it was shown that reachability cannot be maintained with binary quantifier-free programs. 

\shortOrLong{
\begin{proofsketchof}{Theorem \ref{theorem:graphqueries:regular}}

  Let $L$ be a regular path query and let $\calA = (Q, \Sigma, \delta, s, F)$ be a DFA with $L = L(\calA)$. The dynamic program uses a binary relation symbol $R_{p,q}$ for every pair of states $(p, q) \in Q^2$, as well as a binary designated query symbol $R$. The idea is that for a labeled graph $G$, the relation $R_{p,q}$ contains all tuples $(x, y) \in V^2$ such that $\calA$, for some labeled path $\rho$ from $x$ to $y$, can read $\rho$ by starting in state $p$ and ending in state $q$.
  
  The update formulas for the relations $R_{p,q}$ are slightly more involved than the formulas for maintaining reachability under insertions. This is because $\calA$ might reach a state $q$ from a state $p$ only by reading a labeled path from $x$ to $y$ that contains one or more loops. The crucial observation is, however, that for deciding whether $(x,y)$ is in $R_{p,q}$ it suffices to consider paths that contain the node $x$ at most $|Q|$ times (as paths that contain $x$ more than $|Q|$ times can be shortened). This can be checked by quantifier-free update formulas.
\end{proofsketchof}
}{}
\toLongAndAppendix{
\begin{proof}
  The following notion will be useful. Let $\calA$ be a deterministic finite state automaton (short: DFA) and let $G$ be a labeled graph. Then a path $\rho$ in $G$ can be \emph{read by $\calA$ starting in a state $p$ and ending in a state $q$} if $\calA$ can reach state $q$ from state $p$ by reading the label sequence $\lambda(\rho)$ of $\rho$. 
  
  Let $L$ be a regular path query and let $\calA = (Q, \Sigma, \delta, s, F)$ be a DFA with $L = L(\calA)$. We construct a $\DynProp$-program $\prog$ that maintains $L$. 
  
  The program $\prog$ has input schema $\{E_\sigma \mid \sigma \in \Sigma\}$ and an auxiliary schema that contains a binary relation symbol $R_{p,q}$ for every pair $(p, q) \in Q^2$ of states, as well as a binary designated query symbol $R$. The simple idea is that in a state $\state$ with underlying labeled graph $G$, the relation $R^\state_{p,q}$ contains all tuples $(x, y) \in V^2$ such that $\calA$, for some labeled path $\rho$ from $x$ to $y$, can read $\rho$ by starting in state $p$ and ending in state $q$.
  
  The update formulas for the relations $R_{p,q}$ are slightly more involved than the formulas for maintaining reachability under insertions. This is because $\calA$ might reach a state $q$ from a state $p$ only by reading a labeled path from $x$ to $y$ that contains one or more loops. The crucial observation is, however, that for deciding whether $(x,y)$ is in $R_{p,q}$ it suffices to consider paths that contain the node $x$ at most $|Q|$ times (as paths that contain $x$ more than $|Q|$ times can be shortened). This suffices to maintain the relations $R_{p,q}$ dynamically.
  
  The update formulas for $R_{p,q}$ and $R$ are as follows: 
  \begin{align*}
    \uf{R_{p,q}}{\ins_{E_\sigma}}{u,v}{x,y} &\df R_{p,q}(u,v) \vee \bigvee_{p', q'} \Big(R_{p, p'}(x,u) \wedge \varphi^{|Q|}_{p',q'}(u,v) \wedge R_{q', q}(v, y) \Big) \\
    \uf{R}{\ins_{E_\sigma}}{u,v}{x,y} & \df \bigvee_{f \in F} \uf{R_{s,f}}{\ins_{E_\sigma}}{u,v}{x,y}
  \end{align*}
  Here the formula $\varphi^{|Q|}_{p',q'}(u,v)$ shall only be satisfied by tuples $(u,v)$ for which there exists a path $\rho$ from $u$ to $v$ such that $\calA$ can read $\rho$ by starting in $p'$ and ending in $q'$. It shall be satisfied by all such tuples with a witness \mbox{path $\rho$} that contains node $u$ at most $|Q|$ times.
  
  We inductively define, for every $1 \leq i \leq |Q|$ and all $p,q \in Q$, the slightly more general formulas $\varphi^i_{p,q}(u,v)$ as follows:%
  \begin{align*}
    \varphi^1_{p,q}(u,v) & \df [(p, \sigma, q) \in \delta] \vee R_{p,q}(u,v) \\
    \varphi^i_{p,q}(u,v) & \df \varphi^{i-1}_{p,q}(u,v) \vee \bigvee_{p', q'} \Big( \varphi^1_{p,p'}(u,v) \wedge R_{p', q'}(v,u) \wedge \varphi^{i-1}_{q',q}(u,v)\Big)
  \end{align*}
\end{proof}
}

Capturing non-regular path queries by first-order update formulas seems to be significantly harder than capturing CRPQs. We provide only some preliminary results for restricted classes of graphs and modifications.

When all distances for all pairs of nodes can be maintained for a restricted class of graphs, then also non-regular and even non-contextfree path queries can be maintained (e.g. the language $\{a^nb^nc^n\mid n \in  \N\}$). \shortOrLong{Later in Theorem \ref{theorem:distance:insertions} and Theorem~\ref{theorem:distances:undirectedacyclic} we show that distances can be kept up-to-date on acyclic and undirected graphs, as well as on directed graphs under insertions. This implies the following result.}{}
\atheorem{theorem:graphqueries:nonregular}{
  \begin{enumerate}
    \item There is a non-context-free path query that can be maintained in $\DynFO$ on acyclic and undirected $\Sigma$-labeled graphs.
    \item There is a non-context-free path query that can be maintained in $\DynFO$ when only insertions are allowed.
  \end{enumerate}  
}
\toLongAndAppendix{
\begin{proof}
  The non-context-free path query induced by $L = \{a^nb^nc^n\}$ can be maintained since for a graph $G$ distances on $G_a,G_b,G_c$ can be kept up-to-date for those restrictions (see Theorem \ref{theorem:distance:insertions} and Theorem \ref{theorem:distances:undirectedacyclic}). The arithmetic needed for those theorems can be simulated by Proposition~\ref{prop:removing-arithmetic}. We note that also path queries induced by languages such as $\{a^nb^{n+m}c^m\}$ can be maintained by first-order update formulas. %
\end{proof}
}

On acyclic graphs, all context-free path queries can be maintained. It is known that context-free languages are in $\DynFO$ \cite{GeladeMS12} and that the Dyck language with two types of parentheses can be maintained on acyclic graphs \cite{WeberS07}. Generalizing the techniques used for those two results yields the following theorem.

\atheorem{theorem:graphqueries:acycliccontextfree}{
All context-free path queries can be maintained in $\DynFO$ on acyclic graphs.
}

\toMainAndAppendix{To prove Theorem \ref{theorem:graphqueries:acycliccontextfree}, we fix a context-free language $L$ and a grammar \mbox{$\calG = (V,\Sigma, S, P)$} for $L$. We assume, without loss of generality, that $\calG$ is in Chomsky normal form, that is, it has only rules of the form $X \rmapsto YZ$ and~$X \rmapsto \sigma$.  
Furthermore, if $\epsilon \in L$ then $S \rmapsto \epsilon \in P$ and no right-hand side of a rule contains~$S$. 
We write~\mbox{$Z \derive^* w$} if $w \in (\Sigma \cup V)^*$ can be derived from $Z \in V$ using rules of $\calG$.}

\shortOrLong{
The dynamic program maintaining $L$ on acyclic graphs uses $(2k+2)$-ary relation symbols $R_{X \rmapsto Y_1, \ldots, Y_k}$, for $k \in \{1,2,3\}$. The intention is that for an input graph database $G$, the relation $R_{X \rmapsto Y_1, \ldots, Y_k}$ contains a tuple $(x_1, y_1,\ldots, x_{k+1}, y_{k+1})$ if and only if there are strings~\mbox{$s_1, \ldots,  s_{k+1} \in \Sigma^*$} such that $X \derive^* s_1 Y_1 s_2 \cdots s_k Y_k s_{k+1}$ and there is an $s_i$-path $\rho_i$ from $x_i$ to $y_i$ in $G$. 
The paths $\rho_i$ are called \emph{witnesses} for $(x_1, y_1,\ldots, x_{k+1}, y_{k+1}) \in R_{X \rmapsto Y_1, \ldots, Y_k}$.

In a first step we prove that every relation $R_{X \rmapsto Y_1, \ldots, Y_k}$ is first-order definable from the relations $R_{X \rmapsto Y}$, so it actually suffices to only maintain these relations. This proof can be found in the full version of this paper.
}
{}
\toLongAndAppendix{
The dynamic program maintaining $L$ on acyclic graphs will use $4$-ary auxiliary relation symbols $R_{Z \rmapsto Z'}$ for all $Z, Z' \in V$. The intention is that in every \mbox{state $\state$} with input database $G$, the relation $R^\state_{Z \rmapsto Z'}$  contains a tuple $(x_1, y_1,x_2, y_2)$ if and only if there are strings $s_1, s_2 \in \Sigma^*$ such that~\mbox{$Z \derive^* s_1 Z' s_2$} and there is an $s_i$-path $\rho_i$ from $x_i$ to $y_i$ for $i \in \{1,2\}$. The paths $\rho_1$ and $\rho_2$ are called \emph{witnesses} for $(x_1, y_1, x_2, y_2) \in R^\state_{Z \rmapsto Z'}$. Later we will see that whether two nodes are connected by an $L$-path after an update can be easily verified using those relations.

It turns out that for updating the relations $R^\state_{Z \rmapsto Z'}$ it is necessary to have access to $(2k+2)$-ary relations $R^\state_{X \rmapsto Y_1, \ldots, Y_k}$, for $k \in \{1,2,3\}$, which contain a tuple $(x_1, y_1,\ldots, x_{k+1}, y_{k+1})$ if and only if there are strings~\mbox{$s_1, \ldots,  s_{k+1} \in \Sigma^*$} such that $X \derive^* s_1 Y_1 s_2 \ldots s_k Y_k s_{k+1}$ and there is an $s_i$-path $\rho_i$ from $x_i$ to $y_i$ in the input database underlying $\calS$. 

Next, in Lemma \ref{lemma:contextfreeacyclicmult}, we prove that every relation $R^\state_{X \rmapsto Y_1, \ldots, Y_k}$ is first-order definable from the relations $R^\state_{Z \rmapsto Z'}$ (and thus only relations $R^\state_{Z \rmapsto Z'}$ have to be stored as auxiliary data). This lemma is inspired by Lemma 7.3 from \cite{WeberS07}, and its proof is a generalization of the technique used in the proof of  \mbox{Theorem 4.1} in \cite{GeladeMS12}. Afterwards we prove Theorem \ref{theorem:graphqueries:acycliccontextfree}  by showing how to use the relations $R^\state_{Z \rmapsto Z'}$ to maintain $L$ and how to update the relations $R^\state_{Z \rmapsto Z'}$ using the formulas that define relations of the form $R^\state_{X \rmapsto Y_1, Y_2}$ and $R^\state_{X \rmapsto Y_1, Y_2, Y_3}$.

\begin{lemma}\label{lemma:contextfreeacyclicmult}
   For a grammar $\calG$ in Chomsky normal form, $k \geq 2$ and variables $X, Y_1, \ldots, Y_k$ there is a first-order formula $\varphi_{X \rmapsto Y_1, \ldots, Y_k}$ over schema \mbox{$\schema = \{R_{Z \rmapsto Z'} \mid Z, Z' \in V\}$} that defines $R_{X \rmapsto Y_1, \ldots, Y_k}$ in states $\state$ where the relations $R^\state_{Z \rmapsto Z'}$ are as described above. 
\end{lemma}
\begin{proofsketch}
  We explain how $\varphi_{X \rmapsto Y_1, Y_{2}, Y_3}$ tests whether a tuple is contained in $R^\state_{X \rmapsto Y_1, Y_2, Y_3}$. The construction for general $k$ is analogous.  

        \begin{figure}[t!]
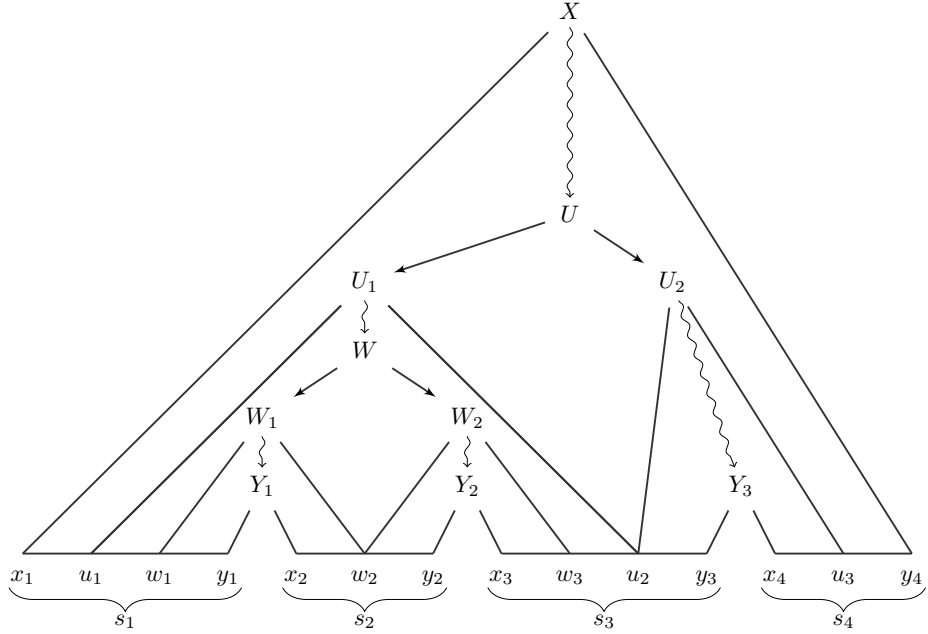
 
      \begin{center}
        \scalebox{0.90}{
          \pictgraphqqueryacyclic
        }
       \caption{Illustration of when a tuple $(x_1, y_1, x_2, y_2, x_3, y_3, x_4, y_4)$ is contained in $R_{X \rmapsto Y_1, Y_2, Y_3}$ in Lemma \ref{lemma:contextfreeacyclicmult}. \label{picture:lemma:contextfreeacyclicmult}}
      \end{center}\vspace{-7mm}
    \end{figure}

  If a tuple $(x_1, y_1, x_2, y_2, x_3, y_3, x_4, y_4)$ is contained in $R^\state_{X \rmapsto Y_1, Y_2, Y_3}$ witnessed by $s_i$-paths $\rho_i$ from $x_i$ to $y_i$ such that $X \derive^* s_1 Y_1 s_2 Y_2 s_3 Y_3 s_4$, then in the derivation tree of $s_1 Y_1 s_2 Y_2 s_3 Y_3 s_4$ from  $X$ there is  a variable $U$ such that $U \rmapsto U_1 U_2$ and either (1) $Y_1$ and $Y_2$ are derived from $U_1$, and $Y_3$ is derived from $U_2$; or (2) $Y_1$ is derived from $U_1$, and $Y_2$ and $Y_3$ are derived from $U_2$. In case (1),  the derivation subtree starting from $U_1$ contains a variable $W$ such that $W \rmapsto W_1 W_2$ and $Y_1$ is derived from $W_1$ and $Y_2$ is derived from $W_2$. Analogously for case (2).  The derivation tree of $X$ for case (1) is illustrated in Figure \ref{picture:lemma:contextfreeacyclicmult}.

  The formula $\varphi_{X \rmapsto Y_1, Y_{2}, Y_3}$ is the disjunction of formulas $\psi_1$ and $\psi_2$, responsible for dealing with the cases (1) and (2) respectively. We only \mbox{exhibit $\psi_1$}, the formula $\psi_2$ can be constructed analogously.  The formula $\psi_1$ guesses the variables $U, U_1, U_2, W, W_1$ and $W_2$, and the start and end positions of strings derived from those variables. Whether $(x_1, y_1, x_2, y_2, x_3, y_3, x_4, y_4)$ is contained in $R^\state_{X \rmapsto Y_1, Y_2, Y_3}$ can then be tested using the relations $R_{Z \rmapsto Z'}$. For simplicity the formula $\psi_1$ reuses the element names $x_i$ and $y_i$ as variable names and is defined as follows:
  \begin{align*}
    \psi_1(x_1, y_1, \ldots, x_4, y_4) =& \exists u_1 \exists u_2  \exists u_3 \bigvee_{\substack{U, U_1, U_2 \in V\\ U \rmapsto U_1U_2 \in P}} \exists w_1  \exists w_2 \exists w_3  \bigvee_{\substack{W, W_1, W_2 \in V\\ W \rmapsto W_1W_2 \in P}} \\
    &\Big( R_{X \rmapsto U}(x_1, u_1, u_3, y_4) \wedge R_{U_1 \rmapsto W}(u_1, w_1, w_3, u_2) \\
    &\wedge R_{W_1 \rmapsto Y_1}(w_1, y_1, x_2, w_2) \wedge R_{W_2 \rmapsto Y_2}(w_2, y_2, x_3, w_3) \\
    & \wedge R_{U_2 \rmapsto Y_3}(u_2, y_3, x_4, u_3)\Big)
  \end{align*}
\end{proofsketch}
 
We now use the relations $R_{Z \rmapsto Z'}$ and the formulas $\varphi_{X \rmapsto Y_1, Y_2, Y_{3}}$ for maintaining context-free path queries on acyclic graphs.

}  
 
\shortOrLong{
\begin{proofideaof}{Theorem \ref{theorem:graphqueries:acycliccontextfree}}
Let $L$ be an arbitrary context-free language and let $\calG = (V,\Sigma, S, P)$ be a grammar for $L$ in Chomsky normal form. We provide a $\DynFO$-program \mbox{$\prog$} with designated binary query symbol $Q$ that \mbox{maintains $L$} on acyclic graphs. The input schema is $\{E_\sigma \mid \sigma \in \Sigma\}$ and the auxiliary schema \mbox{is $\auxSchema = \{R_{X \rmapsto Y} \mid X, Y \in V \} \cup \{T\}$}. The intention of the auxiliary relation symbols $R_{X \rmapsto Y}$ has already been explained above; the relation symbol $T$ shall store the transitive closure of the input graph (where the input graph is the union of all $E_\sigma$).  

As already stated above, the query relation $Q$ is first-order definable from the relations~$R_{X \rmapsto Y}$, as a tuple $(x,y)$ is in the query relation if and only if $x=y$ and $\epsilon \in L$, or there is a $\tau$-labeled edge $(z_1,z_2)$ such that $(x,z_1,z_2,y) \in R_{S \rmapsto U}$ for some $U \in V$ with $U \rmapsto \tau$. 

  It remains to present update formulas for each $R_{X \rmapsto Y}$. After inserting a $\sigma$-edge $(u,v)$, a tuple $(x_1, y_1, x_2, y_2)$ is contained in~$R_{X \rmapsto Y}$ if there are two witness paths $\rho_1$ and $\rho_2$ such that (1) $\rho_1$ and~$\rho_2$ have already been witnesses before the insertion, or (2) only $\rho_1$ uses the new $\sigma$-edge, or (3) only $\rho_2$ uses the new $\sigma$-edge, or (4) both $\rho_1$ and $\rho_2$ use the new $\sigma$-edge. In case (2) the path $\rho_1$ can be split into a path from $x_1$ to $u$, the edge $(u,v)$ and a path from $v$ to $y_1$. Similarly in the other cases and \mbox{for $\rho_2$}. 
This can be expressed using the first-order formulas defining $R_{X \rmapsto Y_1, \ldots, Y_k}$.

After deleting  a $\sigma$-edge $(u,v)$ a tuple $(x_1, y_1, x_2, y_2)$ is in $R_{X \rmapsto Y}$ if it still has witness paths $\rho_1$ and $\rho_2$  from $x_1$ to $y_1$ and from $x_2$ to $y_2$, respectively. The update formula for  $R_{X \rmapsto Y}$ verifies that such witness paths exist. Therefore, similar to Example \ref{example:acyclictc}, the formula distinguishes for each $i \in \{1, 2\}$ whether (1) there was no path from $x_i$ to $y_i$ via $(u,v)$ before deleting the $\sigma$-edge $(u,v)$, or (2) there was a path from $x_i$ to $y_i$ via $(u,v)$. 
  
  In case (1) all paths present from $x_i$ to $y_i$ before the deletion of the $\sigma$-edge $(u,v)$ are also present after the deletion. In particular the set of possible witnesses $\rho_i$ remains the same. For case (2), the update formula has to check that there is still a witness path $\rho_i$. Such a path $\rho_i$ has the options (a) to still use the edge $(u,v)$ but for a $\tau \neq \sigma$, and (b) to not use the edge $(u,v)$ at all. 
   
Whether some witness path uses $(u,v)$ can be checked using the relation $T$. Existence of alternative witness paths can be verified similar to Example \ref{example:acyclictc} with the relations $T$ and $R_{X \rmapsto Y_1, \ldots, Y_k}$.
The complete update formulas can be found in the full version. 
\end{proofideaof}
}{}

\toLongAndAppendix{
\begin{proofideaof}{Theorem \ref{theorem:graphqueries:acycliccontextfree}}
Let $L$ be an arbitrary context-free language and let $\calG = (V,\Sigma, S, P)$ be a grammar for $L$ in Chomsky normal form. We provide a $\DynFO$-program \mbox{$\prog$} with designated binary query symbol $Q$ that \mbox{maintains $L$} on acyclic graphs. The input schema is $\{E_\sigma \mid \sigma \in \Sigma\}$ and the auxiliary schema \mbox{is $\auxSchema = \{R_{X \rmapsto Y} \mid X, Y \in V \} \cup \{T\}$}. The intention of the auxiliary relation symbols $R_{X \rmapsto Y}$ has already been explained above; the relation symbol $T$ shall store the transitive closure of the input graph (where the input graph is the union of all $E_\sigma$).  

Before showing how to update the relations $R_{X \rmapsto Y}$, we state the update formulas for the query relation $Q$. The update formulas distinguish whether the  witness path is of length 0
  or of length at least 1.  The updated relations~$R_{X \rmapsto Y}$ are used for the latter case.  
  \begin{align*}
     \uf{Q}{\ins_{E_\sigma}}{u, v}{&x,y} \df  ([S \rmapsto \epsilon \in P] \wedge x = y ) \\
     & \vee \exists z_1 \exists z_2 \bigvee_{\substack{U \in V \\ U \rmapsto \tau \in P}} \big(\uf{R_{S \rmapsto U}}{\ins_{E_\sigma}}{u,v}{x, z_1, z_2, y} \wedge E_{\tau}(z_1, z_2) \big)\\
     & \qquad \quad \vee \bigvee_{\substack{U \in V \\ U \rmapsto \sigma \in P}} \big(\uf{R_{S \rmapsto U}}{\ins_{E_\sigma}}{u,v}{x, u, v, y} \big)\\
          \uf{Q}{\del_{E_\sigma}}{u, v}{&x,y} \df  ([S \rmapsto \epsilon \in P] \wedge x = y ) \\
     & \vee \exists z_1 \exists z_2 \bigvee_{\substack{U \in V \\ \tau \neq \sigma \\ U \rmapsto \tau \in P}} \big(\uf{R_{S \rmapsto U}}{\del_{E_\sigma}}{u,v}{x, z_1, z_2, y} \wedge E_{\tau}(z_1, z_2) \big)\\
     & \vee \exists z_1 \exists z_2 \bigvee_{\substack{U \in V  \\ U \rmapsto \sigma \in P}} \big(\uf{R_{S \rmapsto U}}{\del_{E_\sigma}}{u,v}{x, z_1, z_2, y} \wedge E_{\sigma}(z_1, z_2) \wedge (z_1 \neq u \vee z_2 \neq v) \big)\\
  \end{align*}
  
  It remains to present update formulas for each $R_{X \rmapsto Y}$. For simplicity we identify names of variable and elements. 
  
  After inserting a $\sigma$-edge $(u,v)$, a tuple $(x_1, y_1, x_2, y_2)$ is contained in~$R_{X \rmapsto Y}$ if there are two witness paths $\rho_1$ and $\rho_2$ such that (1) $\rho_1$ and~$\rho_2$ have already been witnesses before the insertion, or (2) only $\rho_1$ uses the new $\sigma$-edge, or (3) only $\rho_2$ uses the new $\sigma$-edge, or (4) both $\rho_1$ and $\rho_2$ use the new $\sigma$-edge. In case (2) the path $\rho_1$ can be split into a path from $x_1$ to $u$, the edge $(u,v)$ and a path from $v$ to $y_1$. Similarly in the other cases and \mbox{for $\rho_2$}. Using the formulas from \mbox{Lemma \ref{lemma:contextfreeacyclicmult}} this can be expressed as follows:
  \begin{align*}
     \uf{R_{X \rmapsto Y}}{\ins_{E_\sigma}}{u,v}{x_1, y_1, x_2, y_2} \df & R_{X \rmapsto Y}(x_1, y_1, x_2, y_2) \vee \hspace{2cm} & (1)\\*
      & \bigvee_{\substack{U_1, U_2 \in V\\ U_1 \rmapsto \sigma \in P \\ U_2 \rmapsto \sigma \in P}} \big(\varphi_{X \rmapsto U_1,Y}(x_1,u,v,y_1,x_2, y_2) & (2)\\* 
      & \hspace{1cm} \vee \varphi_{X \rmapsto Y,U_2}(x_1,y_1,x_2,u,v, y_2) & (3)\\*
      & \hspace{1cm} \vee \varphi_{X \rmapsto U_1,Y,U_2}(x_1,u,v,y_1,x_2,u,v, y_2)\big)& (4)
  \end{align*}
  
  After deleting  a $\sigma$-edge $(u,v)$ a tuple $(x_1, y_1, x_2, y_2)$ is in $R_{X \rmapsto Y}$ if it still has witness paths $\rho_1$ and $\rho_2$  from $x_1$ to $y_1$ and from $x_2$ to $y_2$, respectively. The update formula for  $R_{X \rmapsto Y}$ verifies that such witness paths exist. Therefore, similar to Example \ref{example:acyclictc}, the formula distinguishes for each $i \in \{1, 2\}$ whether (1) there was no path from $x_i$ to $y_i$ via $(u,v)$ before deleting the $\sigma$-edge $(u,v)$, or (2) there was a path from $x_i$ to $y_i$ via $(u,v)$. See \mbox{Figure \ref{picture:lemma:contextfreeacyclicb}} for an illustration.

    \begin{figure}[t]
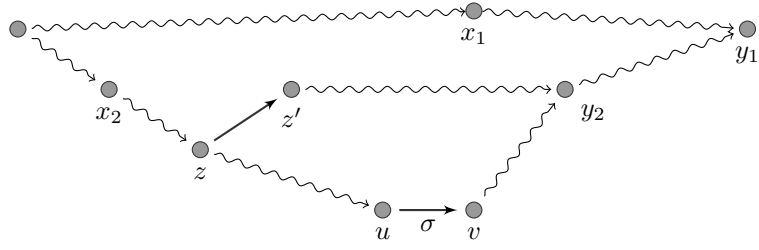
 
      \begin{center}
        \pictgraphqqueryacyclicb
        \caption{Illustration of the update of $R_{X \rmapsto Y}$ after deletion of  $\sigma$-edge $(u,v)$ in the proof of Lemma \ref{lemma:contextfreeacyclicmult}. The nodes $x_1$ and $y_1$ satisfy Condition~(1), whereas nodes $x_2$ and $y_2$ satisfy Condition~(2). \label{picture:lemma:contextfreeacyclicb}}
      \end{center}\vspace{-7mm}
    \end{figure}
  
  In case (1) all paths present from $x_i$ to $y_i$ before the deletion of the $\sigma$-edge $(u,v)$ are also present after the deletion. In particular the set of possible witnesses $\rho_i$ remains the same. For case (2), the update formula has to check that there is still a witness path $\rho_i$. Such a path $\rho_i$ has the options (a) to still use the edge $(u,v)$ but for a $\tau \neq \sigma$, and (b) to not use the edge $(u,v)$ at all. 
   
  The update formula for $R_{X \rmapsto Y}$ is a disjunction over all those cases for the witnesses for $(x_1, y_1)$ and $(x_2, y_2)$. Instead of presenting formulas for all those cases, we explain the idea for two representative cases. All other cases are analogous. 
  
  We first look at the case where $(x_1, y_1)$ satisfies (1), $(x_2, y_2$) satisfies (2) and there are witness paths $\rho_1$ and  $\rho_2$ where $\rho_2$ satisfies (a). The following formula deals with this case:
  \begin{align*}
    & (\neg T(x_1, u) \vee \neg T(v, y_1)) \wedge T(x_2, u) \wedge T(v, y_2) \\ 
    & \quad \quad\quad \quad\quad \quad\quad \quad \wedge \bigvee_{\substack{\tau \neq \sigma, U_2 \in V\\ U_2 \rmapsto \tau \in P}} \big(\varphi_{X \rmapsto Y, U_2}(x_1,y_1,x_2,u,v,y_2) \wedge E_{\tau}(u,v)\big)
  \end{align*}
  In the first line the premises for this case are checked, in the second line it is verified that $\rho_2$ uses $\tau$-edge $(u,v)$ for $\sigma \neq \tau$.

  Now we consider the case where both $(x_1, y_1)$ as well as $(x_2, y_2)$ \mbox{satisfy (2)}, and where there are witness paths $\rho_1$ and $\rho_2$ where $\rho_1$ \mbox{satisfies (a)} and $\rho_2$ satisfies  (b). The existence of such a path $\rho_1$ can be verified as above. For verifying the existence of such a path $\rho_2$, a path not using $(u,v)$ has to be found. This is achieved by relying on the same technique as for maintaining reachability for acyclic graphs (see Example \ref{example:acyclictc}).  The following formula verifies the existence of such $\rho_1$ and $\rho_2$:
  \begin{align*}
    & T(x_1, u) \wedge T(v, y_1) \wedge T(x_2, u) \wedge T(v, y_2) \\ 
    & \quad\quad \wedge \exists z \exists z' \bigvee_{\substack{\tau \neq \sigma, U_1, U_2 \in V\\ U_1 \rmapsto \tau  \in P\\ U_2 \rmapsto \tau' \in P}}\Big(\varphi_{X \rmapsto U_1,Y,U_2}(x_1,u,v,y_1,x_2, z, z',y_2) \wedge E_\tau(u,v) \\
    & \quad\quad\quad\quad\quad\quad\quad\quad\quad\quad\quad\quad  \wedge \big(T(x_2,z) \wedge E_{\tau'}(z,z') \wedge (z \neq u \vee z' \neq v)  \\
    & \quad\quad\quad\quad\quad\quad\quad\quad\quad\quad\quad\quad\quad\quad \wedge T(z',y_2) \wedge T(z,u) \wedge \neg T(z',u)\big) \Big)
  \end{align*}

  Again, in the first line the premises for this case are checked. In the second line $z$ and $z'$ are chosen with the purpose to find an alternative \mbox{path $\rho_2$} (as in Example \ref{example:acyclictc}), and it is verified that $\rho_1$ and $\rho_2$ are witness paths. The third and forth lines verify that $z$ and $z'$ yield an alternative path.%
\end{proofideaof}
}

 \section{Dynamic Extended Conjunctive Regular Path Queries}\label{section:beyondpathqueries}
 \toAppendix{\section{Proofs for Section \ref{section:beyondpathqueries}}}
\makeatletter{}%

In this section we explore the maintainability of ECRPQs. In contrast to path queries, ECRPQs allow for testing properties of tuples of paths between pairs of nodes. Comparing the length of two paths is one of the simplest such properties and is therefore studied first. Afterwards we extend some of the techniques developed for maintaining the lengths of paths to ECRPQs.

\subsection{Maintaining Distances}

Maintaining all distances in arbitrary graphs is one of the big challenges of dynamic complexity. Recall that for maintaining all distances up to $n^c$ a dynamic program has to update, for a graph $G$, a relation $D$ that contains all tuples $(x,y,\ell)$ such that there is a path from $x$ to $y$ of length $\ell$ in $G$, for $0 \leq \ell \leq n^c$.

The 
recent dynamic algorithm for maintaining reachability (see \cite{DattaKMSZ15}) does, unfortunately, not offer hints at how to maintain distances. A dynamic upper bound for distances is provided by Hesse's \DynTCZ-program for reachability \cite{hesse2003tc0}. The program actually maintains the number of different paths of length $\ell$ between every pair of nodes, for any length $\ell$ up to the size of the graph, and thus all distances for all pairs of nodes. The program can be easily modified to compute all distances up to fixed polynomials. 

Here we present preliminary results for maintaining all distances with first-order formulas for restricted modifications as well as for restricted classes of graphs. Furthermore we show that distances cannot be maintained with quantifier-free update formulas.

The \emph{shortest} distance between every pair of nodes can be easily maintained in $\DynFO$ when edges can only be inserted; basically because shortest paths do not contain loops. Maintaining \emph{all} distances for all pairs of nodes under insertions requires some work. 

\begin{theorem}\label{theorem:distance:insertions}
 All distances up to $p(n)$ can be maintained in \DynFO under insertions for every fixed polynomial $p(n)$.
\end{theorem}
\begin{proof}
  We describe how to maintain distances up to $n$; the generalization to distances up to $p(n)$ is straightforward and sketched at the end of the proof. The idea is to maintain a 4-ary relation $A$ that contains a tuple $(x,y,t,\ell)$ if there are $t$ (not necessarily distinct) paths from $x$ to $y$ such that the sum of their lengths is $\ell$. 

    There is a path of length $\ell$ from node $x$ to node $y$ if and only if  $(x,y,1,\ell)$ holds. For maintaining this information, we need the full relation: a path from $x$ to $y$ can use a newly inserted edge $(u,v)$ several times if cycles are present. Also, the path can use an arbitrary combination of cycles including that edge, and each cycle can be used arbitrarily often.

  When inserting an edge $(u,v)$ the updated relation $A$ is defined by the following formula:
    \begin{multline*}
      \uf{A}{\ins_{E}}{u,v}{x,y,t,\ell} \df \exists t_- \exists t_+ \exists t_{\circlearrowleft} \exists \ell_- \exists \ell_{+_1} \exists \ell_{+_2} \exists \ell_{\circlearrowleft} \\
        \Big( A(x,y, t_-, \ell_-)
      \wedge A(x,u,t_+,\ell_{+_1}) \wedge A(v,y,t_+,\ell_{+_2}) \wedge A(v,u,t_{\circlearrowleft},\ell_{\circlearrowleft})\\
        \wedge (t_+=0 \rightarrow t_{\circlearrowleft}=0)
      \wedge t_-+t_+ = t \wedge \ell_-+\ell_{+_1}+\ell_{+_2}+\ell_{\circlearrowleft}
      + t_+ + t_{\circlearrowleft}=\ell \Big)
      \end{multline*}
      
        If there are $t$ paths with total length $\ell$ from $x$ to $y$ after the edge $(u,v)$ is inserted, these paths can be divided into $t_-$ paths that do not use the new edge $(u,v)$, with a total length of $\ell_-$, and $t_+$ paths that use the edge $(u,v)$. 
  Each one of these $t_+$ paths is composed of (i) one path from $x$ to $u$ that does not use $(u,v)$, (ii) the edge $(u,v)$, (iii) possibly some cycles from $v$ back to $v$ created by combining an old path from $v$ to $u$ and the new edge $(u,v)$, and (iv) one path from $v$ to $y$ that does not use $(u,v)$.

  Without considering the cycles in $v$ that use $(u,v)$, in total there are $t_+$ paths from $x$ to $u$ (with total length $\ell_{+_1}$), $t_+$ paths from $v$ to $y$ (with total length $\ell_{+_2}$) and $t_+$ times the new edge $(u,v)$.
  So these paths have total length $\ell_{+_1} + \ell_{+_2} + t_+$. 
  Additionally, let $t_{\circlearrowleft}$ be the number of times the edge $(u,v)$ is used in cycles from $v$ to $v$ in all $t_+$ paths together.
  These cycles can be obtained from $t_{\circlearrowleft}$ paths from $v$ to $u$ of total length $l_{\circlearrowleft}$ and $t_{\circlearrowleft}$ times the new edge $(u,v)$.
  So in total, the $t_+$ paths have a total length of $\ell_{+_1}+\ell_{+_2}+t_++\ell_{\circlearrowleft}+t_{\circlearrowleft}$.
  
  For maintaining distances upto $p(n)$, numbers of this magnitude are encoded by tuples of elements. Arithmetic upto $p(n)$ can be easily defined in a first-order fashion from the built-in arithmetic upto $n$. The above construction then translates in a straightforward way.
\end{proof}

Next we show that all distances for all pairs of nodes in undirected and acyclic graphs can be updated using first-order update formulas. For undirected graphs this slightly extends a result by Grädel and Siebertz \cite{GraedelS12} that the shortest distance can be maintained for undirected paths. For acyclic graphs the maintenance of all distances is a straight-forward extension of the dynamic program for maintaining reachability shown in Example \ref{example:acyclictc}. 

\atheorem{theorem:distances:undirectedacyclic}{
  All distances up to $p(n)$ can be maintained in \DynFO for every fixed polynomial $p(n)$ for (a) undirected graphs, and (b) acyclic graphs. 
}
\shortOrLong{The proof can be found in the full version. %
}{}

\toLongAndAppendix{
\begin{proof}
Again we describe how to maintain distances upto $n$ only; the generalization to distances upto $p(n)$ is straightforward.
\begin{proofenum}
\item We use the simple observation that if two nodes in an undirected graph $G$ are connected by a path of length $m > 0$, then they are connected by a path of length $m+2$ as well, since any edge of the path can be traversed repeatedly. A consequence is that all possible distances between two nodes $x$ and $y$ in an undirected graph can be easily determined if the shortest lengths $d_o$ and $d_e$ of paths of odd and even length between $x$ and $y$ are known: there is a path of length $m$ if $m$ is odd and $m \geq d_o$ or if $m$ is even and $m \geq d_e$ (and if $x=y$, then $x$ is no isolated node).
Thus in order to maintain whether two nodes $x$ and $y$ are connected by a path of length $m$, it suffices (1) to maintain $d_o$ and $d_e$ and (2) to know whether $m$ is even or odd. 

The second part is easy since arithmetic is available.
Maintaining $d_o$ and $d_e$ can be done by maintaining the shortest distances of pairs of nodes in the graph $G \times K_2$, where $K_2$ is the 
complete graphs on nodes $\{1,2\}$. The shortest distance between $(u,1)$ and $(v,1)$ in $G \times K_2$ equals the length of the shortest even path from $u$ to $v$ in $G$, whereas the distance between $(u,1)$ and $(v,2)$ is equal to the length of the shortest odd one. Observe that an edge modification in $G$ spans only two modifications in $G \times K_2$. Since shortest distances in an undirected graphs can be maintained in $\DynFO$~\cite{GraedelS12}, the result follows.	 

 \item This is a simple adaption of the maintenance procedure for the transitive closure of acyclic graphs (see Example \ref{example:acyclictc}). In addition to the transitive closure relation $T$, the dynamic program for distances in acyclic graphs maintains a ternary relation $D$ that contains a tuple $(x,y,\ell)$ if and only if there is a path from $x$ to $y$ of length $\ell$. The update formulas from Example \ref{example:acyclictc} can be adapted easily by using the built-in arithmetic. 
 \begin{align*}
    \uf{D}{\ins_E}{u,v}{x,y,\ell} \df & \; D(x,y, \ell) \vee \exists d \exists d' \big(d+d'+1=\ell\\
    & \quad\quad\quad\quad\quad\quad\quad\quad\quad\wedge D(x, u, d) \wedge D(v,y, d')\big)\\
    \uf{D}{\del_E}{u,v}{x,y,\ell} \df & \; T(x,y) \wedge \Big(\big((\neg T(x,u) \vee \neg T(v,y)) \wedge D(x,y,\ell)\big) \\
    & \quad \vee \exists z \exists z' \exists d \exists d'\big(d+ d'+1= \ell \wedge D(x,z, d) \wedge E(z,z') \\
    & \quad \quad \wedge (z \neq u \vee z' \neq v) \wedge D(z',y, d') \\
    & \quad \quad \wedge T(z,u) \wedge \neg T(z',u) \big)\Big)
 \end{align*}
\end{proofenum}
\end{proof}
}

In the rest of this subsection we discuss why distance information cannot be maintained by quantifier-free update formulas. So far the goal, when maintaining distances, was to store tuples $(a, b, \ell)$ in some relation if there is a path from $a$ to $b$ of length $\ell$, where the length $\ell$ referred to the built-in arithmetic. It can be easily seen that maintaining distances in this fashion is not possible with quantifier-free formulas (basically because a quantifier-free formula only has access to the numbers represented by the modified nodes). 

Another way of maintaining distance information is to store a $4$-relation that contains a tuple $(a_1, a_2, b_1, b_2)$ if and only if there are paths from $a_1$ to $a_2$ and from $b_1$ to $b_2$ of equal length. We show that this relation cannot be maintained by quantifier-free programs. %

Denote by $\eqlength$ the query on (unlabeled) graphs that selects all tuples $(a_1, a_2, b_1, b_2)$ such that there are paths from $a_1$ to $a_2$ and from $b_1$ to $b_2$ of equal length.

\atheorem{theorem:distance:quantifier-freelower}{
  The query \eqlength cannot be maintained by quantifier-free update formulas, even when the auxiliary relations can be initialized arbitrarily. In particular, ECRPQs and reachability in product graphs cannot be maintained in this setting either.
}

Intuitively this is not very surprising. It is well known that non-regular languages and therefore, in particular, the language $\{a^nb^n \mid n \in \N\}$ cannot be maintained by a quantifier-free program \cite{GeladeMS12}. Thus maintaining whether two isolated paths have the same length should not be possible either. Technical issues arise from the fact that the query \eqlength is over graphs, not strings. Yet the techniques used for proving lower bounds for languages can be adapted.

\toLongAndAppendix{
We employ the following Substructure Lemma from \cite[Lemma 4.1]{ZeumeS15reach} which is a slight variation of Lemma 1 \mbox{from \cite{GeladeMS12}}.

The intuition of the Substructure Lemma is as follows. When updating an auxiliary tuple $\vec c$ after an insertion or deletion of a tuple $\vec d$, a quantifier-free update formula has access to $\vec c$, $\vec d$, and the constants only. Thus if a sequence of modifications changes only tuples from a substructure $\calA$ of $\calS$, then the auxiliary data of $\calA$ is not affected by information outside $\calA$. In particular, two isomorphic substructures $\calA$ and $\calB$ remain isomorphic, when corresponding modifications are applied to them.

The notion of corresponding modifications is formalized as follows. Let~$\pi$ be an isomorphism from a structure $\calA$ to a structure $\calB$. Two modifications~$\delta(\vec a)$ on $\calA$ and $\delta'(\vec b)$ on $\calB$ are said to be \textit{$\pi$-respecting} if $\delta = \delta'$ and~\mbox{$\vec b = \pi(\vec a)$}. Two sequences $\alpha=\delta_1\cdots\delta_m$ and $\beta=\delta'_1\cdots\delta'_m$ of modifications respect $\pi$ if $\delta_i$ and $\delta'_i$ are $\pi$-respecting for every $i\le m$. Recall that~$\updateState{P}{\alpha}{\calS}$ denotes the state obtained by executing the dynamic program $\prog$ for the modification sequence $\alpha$ from state $\calS$.

\begin{lemma}[Substructure Lemma \cite{GeladeMS12}]\label{lemma:quantifierfree:substructure}
  Let $\prog$ be a  \DynProp-program and let $\calS$ and $\calT$ be states of $\prog$ with domains $S$ and $T$. Further let $A \subseteq S$ and $B \subseteq T$ such that $\restrict{\calS}{A}$ and $\restrict{\calT}{B}$ are isomorphic via $\pi$. Then $\restrict{\updateState{P}{\alpha}{\calS}}{A}$ and $\restrict{\updateState{P}{\beta}{\calT}}{B}$ are isomorphic via  $\pi$ for all $\pi$-respecting modification sequences $\alpha$, $\beta$ on $A$ and $B$.
\end{lemma}

\begin{proofof}{Theorem \ref{theorem:distance:quantifier-freelower}}
  Towards a contradiction,  assume that  $\prog = (P, \init, \querys)$ is a dynamic program over schema $\tau = (\inpSchema, \auxSchema)$ that maintains the query \eqlength in its designated query relation $\querys$. Let $n'$ be sufficiently large with respect to $\schema$ and $n$ be sufficiently large with respect to $n'$. Further let $m$ be the highest arity of a relation symbol from $\auxSchema$.

  Let $G = (V, E)$  be the empty graph with $|V| = n$ and let $\state = (V, E, \aux)$ be the state obtained by applying the initialization mapping of $\prog$ to $G$. 

  By Ramsey's Theorem for structures (see, e.g., \cite[Theorem 4.3]{ZeumeS15reach}) and because $n=|V|$ is sufficiently large with respect to $n'$ there is a set $V' \subseteq V$ of size $2n'$ and an order $\norder$ on $V'$ such that all $\norder$-ordered $m$-tuples over $V'$ are of equal atomic $\auxSchema$-type. Let us assume that $V' = A \cup B$ with $A = \{a_1, \ldots, a_{n'}\}$ and $B = \{b_1, \ldots, b_{n'}\}$, and that $a_1 \prec \ldots \prec a_{n'} \prec b_1 \prec \ldots \prec b_{n'}$.

  Let $\state' \df (V, E', \aux')$ be the state of $\prog$  that is reached from $\state$ after inserting the edges $(a_1,a_2), (a_2,a_3), \ldots, (a_{n'-1}, a_{n'})$. 
  
  Our goal is to find $i_1, i_2, i_3$ with $i_1 < i_2 < i_3$ such that the substructures $\restrict{\calS'}{\{a_{i_1}, a_{i_2}, b_1 \ldots, b_{n'}\}}$ and $\restrict{\calS'}{\{a_{i_1}, a_{i_3}, b_1 \ldots, b_{n'}\}}$ are isormorphic. Then, in the state $\state''$ obtained from $\state'$ by inserting the edges $\{(b_1, b_2), (b_2, b_3), \ldots, (b_{i_2-i_1-1}, b_{i_2-i_1})\}$, the tuples $(a_{i_1}, a_{i_2}, b_1, b_{i_2-i_1})$ and $(a_{i_1}, a_{i_3}, b_1, b_{i_2-i_1})$ will either be both in $Q$ or both not in $Q$ (due to the Substructure Lemma). However, there is a path of length $i_2-i_1$ between $a_{i_1}$ and $a_{i_2}$ but not from $a_{i_1}$ to $a_{i_3}$, a contradiction.

  It remains to exhibit such $i_1$, $i_2$ and $i_3$. To this end observe that for all $m$-ary tuples $\vec b_1$ and $\vec b_2$, the tuples $(a_i, a_j, \vec b_1)$ and $(a_i, a_j, \vec b_2)$ have the same atomic type due to the Substructure Lemma. Furthermore, by Ramsey's Theorem for structures, one can find $i_1, i_2, i_3$ such that $(a_{i_1}, a_{i_2}, \vec b_1)$ and $(a_{i_1}, a_{i_3}, \vec b_2)$ have the same atomic type. But then $\calT_1 \df \restrict{\calS'}{\{a_{i_1}, a_{i_2}, b_1 \ldots, b_{n'}\}} \isomorph \restrict{\calS'}{\{a_{i_1}, a_{i_3}, b_1 \ldots, b_{n'}\}}$ via the isomorphism that maps $a_{i_1}$ and each $b_i$ to itself and $a_{i_2}$ to $a_{i_3}$.

\end{proofof}
}

\subsection{Maintaining ECRPQs}

Here we study the maintenance of ECRPQs and provide results in restricted settings.
First we show that answers to an ECRPQ can be maintained in $\DynFO$ on acyclic graphs. Even more, answers to the following extension of ECRPQs introduced in \cite{BarceloLLW12} can still be maintained. An ECRPQ with \emph{linear constraints on the number of occurrences of symbols} on paths over an alphabet  $\Sigma = \{\sigma_1,...,\sigma_k\}$ is of the form 
\[\query(\vec z) \longleftarrow \bigwedge_{1 \leq i \leq m } (x_i, \pi_i, y_i), \bigwedge_{1 \leq j \leq t} R_j(\vec \omega _j),\ A \vec \ell \geq \vec b\]
where $A \in \mathbb Z^{h \times (km)}$ for some $h \in \N$, $\vec b \in \mathbb Z^h$, and $\vec \ell = (\ell_{1,1},...\ell_{1,k},...,\ell_{m,1},...,\ell_{m,k})$. The semantics extends the semantics of ECRPQs as follows: for each $1 \leq i \leq m$ and $1 \leq j \leq k$, 
the variable $\ell_{i,j}$ is interpreted as the number of occurrences of the symbol $\sigma_j$ in the path $\pi_i$. The last clause
of the query $\query$ is true if $A \vec \ell \geq \vec b$ under this interpretation. 

\atheorem{theorem:ecrpqonacyclic}{
  Every ECRPQ with linear constraints on the number of occurrences of symbols is maintainable in \DynFO\ on acyclic graphs.
}
\aproof{
Let $\Sigma = \{\sigma_1,...,\sigma_k\}$. We show how to maintain the answer of an ECRPQ $\query$ with linear constraints with only one regular relation $R$ on an acyclic $\Sigma$-labeled graph $G = (V,E)$. Thus $\query$ is of the form: 
\[\query(\vec z) \longleftarrow \bigwedge_{1 \leq i \leq m} (x_i, \pi_i, y_i), R(\pi_1,\ldots,\pi_m), A \vec \ell \geq \vec b\]
An arbitrary ECRPQ with linear constraints can be rewritten in this form by using closure properties of regular relations. 

In a first step we reduce this problem to a structurally simpler one: the problem of maintaining $\query$ on a $\Sigma$-labeled graph consisting of $m$ disjoint
 acyclic graphs $G_1, \ldots, G_m$, restricted in such a  way that solutions may only map the variables $x_i, y_i$ to nodes in $G_i$, for each $ 1 \leq i \leq m$. 
The simple reduction from the original problem copies the queried graph $m$ times. As $m$ is a constant, this is a bounded first-order reduction.

Let $\calA = (Q,(\Sigma \cup\bot)^m, \delta,s,F)$ be a finite automaton with padding symbol $\bot \not \in \Sigma$ that recognizes the $m$-ary regular relation $R$. The idea is to maintain $(2m+km)$-ary auxiliary relations $R_{p,q}$ for all $p, q \in Q$ intended to store a tuple $(\vec x, \vec y, \vec \ell_1, \ldots, \vec \ell_m)$ with $\vec x = (x_1,\ldots,x_m), \vec y = (y_1,\ldots,y_m)$ and $\vec \ell_i = (\ell_{i,1},\ldots,\ell_{i,k})$ if and only if the state $q$ is reachable from the state $p$ in $\calA$ by reading a tuple of words $(\lambda(\rho_1),\ldots,\lambda(\rho_m))$, where for each $1 \leq i \leq m$,  $\rho_i$ is a path in $G_i$ from $x_i$ to $y_i$, and $\ell_{i,1},\ldots,\ell_{i,k}$ are the number of occurrences of the symbols $\sigma_1,\ldots,\sigma_k$ in the label sequence of $\rho_i$.

We show how to express the query relation $Q$ by these relations. To this end observe that the (fixed) linear inequality system $A \vec \ell \geq \vec b$ can be defined by a $(m\times k)$-ary 
first-order formula $\psi_{A,\vec b}(\vec \ell_1,\ldots,\vec \ell_m)$ that uses the built-in arithmetic. 

The query relation $Q$ is then defined by the following formula:
\[\varphi(\vec z) \df \exists \vec v\; \exists \vec \ell_1 \cdots \exists \vec \ell_m  \bigvee_{f \in F} R_{s,f} (\vec x, \vec y, \vec \ell_1,\ldots,\vec \ell_m) \land \psi_{A,\vec b}(\vec \ell_1,\ldots,\vec \ell_m)\]
Here the existentially quantified variables $\vec v$ correspond to variables of $\query$ that do not occur in the head of the query, and all $x_i$ and $y_i$ occur in either $\vec z$ or $\vec v$.

The update formulas for the relations $R_{p,q}$ are similar in spirit to those for reachability in acyclic graphs used in Example \ref{example:acyclictc}.
}{They are described in detail in the full paper.}{
Suppose an edge $(u,\sigma,v)$ is inserted into the graph $G_i$ for some $i \in \{1, \ldots, m\}$. The update formulas compose runs of $\calA$ from the runs stored in the relations $R_{p,q}$ as follows. For all states $p,q \in Q$, a tuple $(\vec x, \vec y, \vec \ell _ 1,\ldots, \vec \ell _ k)$ shall be in $R_{p,q}$ after the insertion if and only if it was in $R_{p,q}$ before the insertion or if the following conditions are satisfied:
\begin{enumerate}
\item There is a state $p' \in Q$, a tuple of nodes $\vec x' = (x'_1,\ldots,x'_m)$ with $x'_i = u$, and vectors $\vec a_1,\ldots,\vec a_k \in \mathbb N^m$, such that $(\vec x, \vec x', \vec a_1,\ldots,\vec a_k) \in R_{p,p'}$.
\item There is a state $q' \in Q$, a tuple of nodes $\vec y' = (y'_1,\ldots,y'_m)$ with $y'_i = v$, and vectors $\vec b_1,\ldots, \vec b_k \in \mathbb N^m$, such that $(\vec y', \vec y, \vec b_1,\ldots,\vec b_k) \in R_{p',q}$.
\item There is a tuple of symbols  $\vec s \in (\Sigma \cup \bot)^m$ such that 
  \begin{enumerate}
    \item[(i)] $s_i = \sigma$, 
    \item[(ii)] $s_j = \bot$ for each $j \neq i$ with $x'_j = y'_j$, and
    \item[(iii)] there is an edge $(x'_j, s_j, y'_j) \in E_j$ for each $j \neq i$ with $x'_j \neq y'_j$
  \end{enumerate}
  and $\calA$ has a transition from $p'$ to $q'$ by reading $\vec s$.
\item $\vec \ell_j = \vec a_j + \vec b_j + \vec c_{\sigma_j}$ for each $j \in \{1, \ldots, k\}$, where $\vec c_{\sigma_j} \in \{0,1\}^m$ is the vector whose $r$th component is $1$ if the $r$th component of $\vec s$ is $\sigma_j$, and 0 otherwise.
\item $A \vec \ell \geq \vec b$, where $\vec \ell$ is the concatenation of $\vec \ell_1, \ldots, \vec \ell_k$.
\end{enumerate}
The conditions (a)-(c) can be easily expressed by first-order formulas using existential quantification. The conditions (d)-(e) can be expressed by using built-in arithmetic: since the graphs $G_1,\ldots,G_k$ are acyclic, it is easy to see that numbers used in those conditions are polynomial in the size of the active domain. We can therefore build the needed arithmetic incrementally by Proposition~\ref{prop:removing-arithmetic}. 

Deletions can be handled along the same lines by using the technique from  Example~\ref{example:acyclictc}. 
}

It remains open whether the answer relation of ECRPQs can be maintained on general graphs, even when only insertions are allowed. Yet when the rational relations are restricted to be unary, the ECRPQs can be maintained under insertions. More formally, a \emph{CRPQ with linear constraints on the number of occurrences of symbols} over $\Sigma = \{\sigma_1,\ldots,\sigma_k\}$ is of the form
\[\query(\vec z) \longleftarrow \bigwedge_{1 \leq i \leq m } (x_i, \pi_i, y_i), \bigwedge_{1 \leq j \leq m}L_j(\pi_j),\ A \vec \ell \geq \vec b\]
where $L_j$ is a unary rational relation (that is, a regular language), and $A$, $\vec b$ and $\vec \ell$ are as in the definition of ECRPQs with linear constraints.

\atheorem{theorem:ecrpqinsertions}{
  Every CRPQ with linear constraints on the number of occurrences of symbols is maintainable in \DynFO\ under insertions.
}
\aproof{
Let $\Sigma = \{\sigma_1, \ldots, \sigma_k\}$ and $\query$ be a CRPQ over $\Sigma$ with linear constraints on the number of occurrences of symbols as above. Further let $\calA_j = (Q_j,\Sigma,\delta_j,s_j,F_j)$, $1 \leq j \leq m$, be finite state automata for the regular languages $L_j$ occurring in $\query$. 

We exhibit a \DynFO-program with built-in arithmetic for maintaining $Q$ on general graphs under insertions. The necessity for built-in arithmetic can be removed by Proposition~\ref{prop:removing-arithmetic}. 

The idea is similar to the proof of the previous Theorem \ref{theorem:ecrpqonacyclic}. We maintain $(k+2)$-ary auxiliary relations $R^j_{p,q}$ for each $j \in \{1, \ldots, m\}$ and all $p,q \in Q_j$ with the intention that $R^j_{p,q}$ stores a tuple $(x,y,\ell_1,\ldots,\ell_k)$ if and only if the state $q$ is reachable from state $p$ in the automaton $\calA_j$ by reading the label of a path $\rho$ between $x$ and $y$ in $G$ such that $\ell_1,\ldots,\ell_k$ are the number of occurrences of $\sigma_1, \ldots, \sigma_k$ in $\rho$.

Before sketching how to maintain the relations $R^j_{p,q}$, we show how they can be used to express the answer of $\query$. As in the proof of Theorem \ref{theorem:ecrpqonacyclic} the (fixed) linear inequality system $A \vec \ell \geq \vec b$ can be defined by a $(m\times k)$-ary 
first-order formula $\psi_{A,\vec b}(\ell_{1,1},\ldots,\ell_{m,k})$ that uses the built-in arithmetic. Then a tuple $\vec u$ of nodes in $G$
is in the answer of $\query$ if and only if the following formula holds:
\[\varphi(\vec z) \df \exists \vec v\; \exists \ell_{1,1},\ldots,\ell_{m,k}, \bigwedge_{1 \leq j \leq m} \left(\bigvee_{f \in F_j} R^j_{s_j,f} (x_j, y_j, \ell_{j,1},\ldots,\ell_{j,k}) \right ) \land \psi_{A,\vec b}(\ell_{1,1},\ldots,\ell_{m,k})\]
Here the existentially quantified variables $\vec v$ correspond to variables of $\query$ that do not occur in the head of the query, and all $x_j$ and $y_j$ occur in either $\vec v$ or $\vec z$.%

A small technical issue arises from the fact that it is not obvious why the length of paths $\rho_1,\ldots,\rho_m$ witnessing that a tuple of nodes $\vec u$ is in the answer of $\query$ is polynomially bounded. This, however, is necessary for being able to quantify the length $\ell_{1,1},\ldots,\ell_{m,k}$ and to use the built-in arithmetic for computations. Fortunately the length of (shortest) witness paths can be bounded by a fixed polynomial in the size of the active domain.  This has been shown even for ECRPQs with such linear constraints in \cite[Lemma 8.6]{BarceloLLW12}. 

Now we show how to maintain the relations $R^j_{p,q}$. The following notion is useful. A relation $R$ stores the \emph{Parikh distances} of a $\Sigma$-labeled graph if it contains a tuple $(x,y,\ell_1,\ldots\ell_k)$ if and only if there is a path $\rho$ between $x$ and $y$ such that its label $\lambda(\rho_i)$ contains $\ell_i$ occurrences of the symbol $\sigma_i$ for each $1 \leq i \leq m$. We observe that the relations $R^j_{p,q}$ can be defined from the Parikh distance relations of the product graphs $G \times A_j$. Since the automata $A_j$ are fixed, a modification of $G$ yields a bounded number of first-order definable modifications to $G \times A_j$. 

Thus in order to maintain $R^j_{p,q}$, it suffices to be able to maintain the Parikh distance relation of a $\Sigma$-labeled graph under insertions. However, the dynamic program for maintaining distances under insertions from Theorem \ref{theorem:distance:insertions} can be easily generalized to maintain Parikh distances.%
}{}{
For the sake of completeness we present the general construction.  The goal is to maintain an auxiliary relation $S$ intended to store a tuple $(x,y,t, \vec \ell)$ with $\vec \ell \df (\ell_1,\ldots,\ell_k)$ if there are (not necessarily distinct) paths $\rho_1,\ldots,\rho_t$ from $x$ to $y$ in $G$ such that %
each symbol $\sigma_i \in \Sigma$ appears exactly $\ell_i$ times among all $\rho_1,..,\rho_t$ paths. The update formula for $S$ after inserting an edge $(u,\sigma,v)$ is as follows:
  \begin{align*}
      \uf{S}{\ins_{E_{\sigma_i}}}{u,v}{x,y, t,\vec \ell} \df & \; \exists t_- \exists t_+ \exists t_{\circlearrowleft} \exists \vec \ell_- \exists \vec \ell_{+_1} \exists \vec \ell_{+_2} \exists \vec \ell_{\circlearrowleft} \Big( A(x,y, t_-, \vec \ell_-) \\
      & \qquad \wedge A(x,u,t_+,\vec \ell_{+_1}) \wedge A(v,y,t_+,\vec \ell_{+_2}) \\
      & \qquad \wedge A(v,u,t_{\circlearrowleft},\vec \ell_{\circlearrowleft}) \wedge (t_+=0 \rightarrow t_{\circlearrowleft}=0)\\
      & \qquad \wedge t_-+t_+ = t \wedge \vec \ell_-+\vec \ell_{+_1}+\vec \ell_{+_2}+\vec \ell_{\circlearrowleft}+(t_+ + t_{\circlearrowleft})\vec e_i=\vec \ell \Big)
      \end{align*}
Here, for clarity we quantify $k$-ary tuples of variables. The tuple $\vec e_i$ contains zeroes except for its $i$-th component, which is 1. 

The correctness of this update formula follows immediately from the proof of Theorem~\ref{theorem:distance:insertions}.
}

We remark that already boolean ECRPQs cannot be maintained under insertions in $\DynProp$ due to lower bounds for non-regular languages \cite{GeladeMS12}, and boolean CRPQs with $k+2$ existentially quantified node variables cannot be maintained in $\DynProp$ with $k$-ary relations due to a lower bound for the $k$-clique query \cite{ZeumeS14cq}.

  \section{Maintaining Reachability in Product Graphs}\label{section:reachabilityInProductGraphs}
  \toAppendix{\section{Proofs for Section \ref{section:reachabilityInProductGraphs}}}
\makeatletter{}%

In this final section we study the reachability query for product graphs. In addition to its importance for the evaluation of fixed graph queries, reachability in graph products can be used to maintain the result of regular path queries in combined complexity (i.e., when the query is subject to modifications as well). Furthermore it is relevant in model checking, where subsystems correspond to factors in product graphs (see, e.g., \cite{BaierK08}).

The results for maintaining all distances obtained in the previous section immediately transfer to reachability in simple graph products (see the discussion at the end of Section \ref{section:preliminaries}). A small technical obstacle arises from the fact that the reachability query does not come with built-in arithmetic, while the distance query studied so far does. However, this is not a problem due to Proposition~\ref{prop:removing-arithmetic}.

\atheorem{theorem:products:meta}{
  Let $\calG$ be a class of graphs and $m \in \N$. If all distances up to $n^m$ on $\calG$ can be maintained in \DynFO with built-in arithmetic, then reachability in the product of $m$ $\calG$-graphs is maintainable in \DynFO (without built-in arithmetic).
}
\toLongAndAppendix{
\begin{proof}
  For $\DynFO$ with arithmetic this follows from Fact \ref{prop:reach-prod-dists}.  As reachability is a domain independent query the result follows from Proposition~\ref{prop:removing-arithmetic}. 
\end{proof}  
}

Shortest paths in products of acyclic and undirected graphs are of length at most  $n$ and~$n^2$, respectively.  For these two classes of graphs, reachability can therefore be maintained in products of polynomially many factors using the program for all distances. More precisely, this is doable for reachability between two specified nodes $\vec s$ and $\vec t$ as opposed to all pairs of nodes (as there are exponentially many nodes in such product graphs). 

For directed graphs, shortest paths in products of polynomially many graphs can be of exponential length. For this reason, the approach to maintain reachability in such products via distances fails. 
Even more, it is unlikely that there is a $\DynFO$-program for this problem: it could be used to decide reachability in the product of polynomially many graphs in $\PTIME$, which is NP-hard. This follows from a reduction from emptiness of intersections of unary regular expressions which is known to be NP-hard \cite{galil1976hierarchies}. 

\begin{corollary}\label{theorem:products:undirected-and-acyclic-products}
Reachability can be maintained in \DynFO in the product of
\begin{enumerate}
\item polynomially many undirected graphs,
\item polynomially many acyclic graphs, and
\item a constant number of directed graphs under insertions.
\end{enumerate} 
\end{corollary}
This follows immediately from Theorem \ref{theorem:products:meta}, Theorem \ref{theorem:distances:undirectedacyclic} and Theorem \ref{theorem:distance:insertions}. Reachability in products of an undirected and an acyclic graph and similar constellations can, of course, also be maintained. 

For labeled graph products, the following corollary follows immediately from the proof of Theorem \ref{theorem:ecrpqonacyclic}.

\begin{corollary}\label{coro:graphqueries:acyclicproducts}
  Reachability in products of constantly many acyclic $\Sigma$-labeled graphs can be maintained in $\DynFO$. 
\end{corollary}

In the following we generalize Corollary \ref{theorem:products:undirected-and-acyclic-products} to a broader class of graph products. In the product graphs considered so far, there is an edge from a node $(x_1, \ldots, x_m)$ to a node $(y_1, \ldots, y_m)$ if there is an edge $(x_i, y_i)$ in every factor $G_i$. This can be seen as a completely synchronized traversal through the given graphs. The graph products to be introduced next allow for more flexible, partially synchronized traversals.

Let $(G_i)_{1 \leq i \leq m}$ be a sequence of graphs with $G_i \df (V_i, E_i)$, and let $A \df (\vec a_1, \ldots, \vec a_k)$ be a list of tuples from $\{0,1\}^m$, called \emph{transition rules}. We often identify $A$ with the matrix that has the tuples $\vec a_i$ as columns. The \emph{generalized graph product of $(G_i)_i$ with respect to $A$}, denoted $\prod^A_i G_i$, has nodes $V_1 \times \cdots \times V_m$ and edges $(\vec x, \vec y)$ defined by the first-order formula 

$$
 \bigvee_{\stackrel{\vec a \in \calA}{\vec a = (a_1, \ldots, a_m)}} \bigwedge_{a_i = 0} x_i = y_i\  \land\ \bigwedge_{a_i = 1} E_i(x_i,y_i)$$

For example, the usual product of two graphs is defined by the transition rule $\{(1,1)\}$, and the so called {\em cartesian product} is defined by the rules $\{(1,0),(0,1)\}$. We remark that generalized graph products have also been called \emph{non-complete extended p-sums},  short: NEPS (see, for example, \cite{stevanovic1999neps}). %

\atheorem{theorem:graphproducts:neps}{
  Reachability in generalized product graphs is maintainable in \DynFO\ under modifications to factors and transitions rules\footnote{We permit single bit modifications to $A$, that is, modifying one bit of a transition rule at a time.} for 
  \begin{enumerate}
    \item a constant number of directed graphs under insertions and a constant number of transition rules,
    \item polynomially many acyclic graphs and a constant number of transition rules,
    \item polynomially many undirected graphs and polynomially many transition rules. 
  \end{enumerate}
}
\shortOrLong{
\begin{proofsketch}
 For (a) and (b),  the key observation is that reachability in generalized graph products can be reduced to finding a solution of small natural numbers to a linear equation system. For proving (c), a characterization of reachability in generalized products of undirected graphs from \cite{stevanovic1999neps} as well as the dynamic program for maitaining the rank of a matrix from \cite{DattaKMSZ15} is used. Details can be found in the full paper. 
\end{proofsketch}

}{}
\toLongAndAppendix{
\begin{proofsketch}
  As usual we assume built-in arithmetic, which can be removed by Proposition~\ref{prop:removing-arithmetic}.

  For (a) and (b),  the key observation is that reachability in generalized graph products can be reduced to finding a solution in natural numbers to a linear equation system. Let $(G_i)_ {1 \leq i \leq m}$ be a list of graphs, $\vec x = (x_1, \ldots, x_m)$ and $\vec y = (y_1, \ldots, y_m)$ nodes of $\prod^A_i G_i$, and let $$D \df \{\vec d = (d_1, \ldots, d_m) \mid \text{there is a path from $x_i$ to $y_i$ in $G_i$ of length $d_i$, for each $1 \leq i \leq m$}\}.$$ Then there is a path from $\vec x$ to $\vec y$ in $\prod^A_i G_i$
  if and only if there is a tuple $\vec d \in D$ and $n_1, \ldots, n_k \in \N$ such that $ n_1 \vec a_1 + \ldots n_k \vec a_k = \vec d$. A shortest path witnessing that two tuples $\vec x$ and $\vec y$ are connected in a generalized product of constantly many directed graphs (or of polynomially many acyclic graphs) can be of at most polynomial length. In particular, we can restrict numbers $n_1, \ldots, n_k \in \N$ to be of polynomial size. 

  The dynamic program for maintaining reachability in those graph products works as follows. It maintains all distances for each of the factors. Upon modification of a graph $G_i$, the program updates all distances for $G_i$ (using the program for maintaining all distances). Then it guesses $n_1, \ldots, n_k$ by using existential quantification, computes $\vec d \df n_1 \vec a_1 + \ldots n_k \vec a_k$, and checks for each component $d_i$ of $\vec d$ that in $G_i$ there is a path from $x_i$ to $y_i$ of length $d_i$. Modifications of the transition rule are handled in a similar way.
 
  For (c) we rely on the following fact, which is a consequence of the proof of Theorem 2 in~\cite{stevanovic1999neps}.
  
  \begin{fact*}
      Let $G \df \prod^A_i G_i$ be the generalized product of the undirected graphs $(G_i)_{1 \leq i \leq m}$ with respect to a list $A$ of $k$ transition rules. Let $\vec x \df (x_1, \ldots, x_m)$ and $\vec y = (y_1, \ldots, y_m)$ be two nodes of $G$, let $C_i$ be the connected component of $x_i$ in $G_i$ and assume that $C_{i_1}, \ldots, C_{i_\ell}$ are the only bipartite components. Then there is a path from $\vec x$ to $\vec y$ in $G$ if and only if
      \begin{itemize}
	\item for each $i \in \{1, \ldots, m\}$ there is a path from $x_i$ to $y_i$ in $G_i$, and
	\item the linear equation system $B \vec x = \vec d$ is solvable over $\Z_2$ where 
	\begin{itemize}
	  \item $B$ is obtained from $A$ by setting rows $r \notin \{i_1, \ldots, i_\ell\}$ to zero, and
	  \item the $r$th component of $\vec d \in \Z_2^\ell$ is the parity of the distances between $x_{r}$ and $y_{r}$ for $r \in \{i_1, \ldots, i_\ell\}$ and zero for $r \notin \{i_1, \ldots, i_\ell\}$.
	\end{itemize}
      \end{itemize}
  \end{fact*}
  Note that since the component of $x_r$ with $r \in \{i_1, \ldots, i_\ell\}$ is bipartite, all paths between $x_{r}$ and $y_{r}$ have the same parity.

  We use the above fact to construct a $\DynFO$-program that maintains whether there is a path from $\vec x$ to $\vec y$ in the generalized product of polynomially many undirected graphs $(G_i)_i$ with respect to polynomially many transition rules $A$ under single edge modifications to factors and single bit modifications to transition rules. 
  
  The dynamic program maintains auxiliary data that contains (1) all distances for each of the factors (and thus, in particular, also whether there is a path from $x_i$ to $y_i$ and whether the component $C_i$ that contains $x_i$ is bipartite) and (2) whether the equation system $B\vec x = \vec d$ has a solution over $\Z_2$. For the latter the program maintains whether $\rank(B) = \rank(B, \vec d)$ over $\Z_2$. 
  
  It is known that the rank of matrices can be maintained in $\DynFO$ \cite{DattaKMSZ15}. Even more, as observed by William Hesse, the algorithm from \cite{DattaKMSZ15} can maintain the rank even when whole rows may be replaced. 

  Upon modification of a factor $G_i$, the program updates all distances for $G_i$ using the dynamic program for maintaining distances in undirected graphs. If the modification yields a bipartite component $C_i$ of $G_i$, then the $i$th row of $B$ is replaced by the $i$th row of $A$ and $d_i$ is set to the parity of paths between $x_i$ and $y_i$. If the component $C_i$ became non-bipartite, then the $i$th row of $B$ is replaced by the all-zero row of $A$ and $d_i$ is set to zero.
   On the other hand, if the $i$-th bit of transition rule $\vec a_j$ in $A$ is modified, then the $i$-th row of $B$  is modified only at its $j$-th entry if and only if $C_i$ is bipartite. 
   In all scenarios, at most one row of $B$ is modified.  The program can therefore maintain the ranks of $B$ and $(B, \vec d)$ accordingly.
  Finally, if $y_i$ is reachable from $x_i$ in $G_i$ for every  $1 \leq i \leq m$ and  $\rank(B) = \rank(B, \vec d)$ then the query bit of the dynamic program is set true. 

  The update operations described above can be expressed by first-order formulas with the aforementioned auxiliary data.

\end{proofsketch}
}
Observe that deciding reachability in generalized products of (1) polynomially many graphs with constant many transition rules and of (2) polynomially many acyclic graphs with polynomially many transitions rules are $\NP$-hard problems. More precisely, the first generalizes reachability in the product of polynomially many graphs, which we already discussed above. As for
the second, notice that the problem of deciding the existence of a 0-1 solution of a linear equation $A\vec x = \vec 1$, which is known to be NP-hard even for a 0-1 matrix $A$ \cite[Chapter 8]{dasgupta2006algorithms}, can be straightforwardly reduced to  reachability in the generalized product of acyclic graphs when polynomially many transition rules are allowed (by using the distance and linear equations characterization used in the proof of Theorem \ref{theorem:graphproducts:neps}). These problems
are thus unlikely to be maintainable in \DynFO.

%
%
%
%
%
%
%
%
%
%
%
%
%
%
%
%
%
%
%

%
%
%
%
%
%
%
%
%
%
%
%
%
%
%
%
%
%
%
%
%
%
%
%
%
%
%
%
%
%
%
%
%
%
%
%
%
%
%
%
%
%
%
%
%
%
%
%
%
%
%
%
%
%
%
%
%
%
%
%
%
%
%
%
%
%
%
%
%
%
%
%
%
%
%
%
%
%
%
%
%
%
%
%
%
%
%
%
%
%
%
%
%
%
%
%
%
%
%
%

  \section{Conclusion}\label{section:conclusion}
\makeatletter{}%
In this article we explored graph query languages in the dynamic descriptive complexity framework introduced independently by Dong, Su and Topor, and Patnaik and Immerman. Furthermore we investigated the strongly related question, under which conditions distances in graphs as well as reachability in product graphs can be maintained.  Our work is only a first step towards a systematic understanding of graph queries in dynamic graph databases. In the following we discuss some interesting directions for further research. 

For several restricted classes of graphs we exhibited first-order update programs for maintaining distances. We also showed that quantifier-free update formulas do not suffice. It remains open, whether distances can be maintained for general graphs; we conjecture that this is the case.
\begin{openproblem}
  Exhibit a $\DynFO$-program for maintaining distances.
\end{openproblem}

As we have seen, reachability in products of labeled graphs is related to maintaining fragments of the graph query language ECRPQ. While we showed that reachability can be maintained in labeled products of acyclic graphs, this problem is already much harder for products of undirected, labeled paths---not to mention arbitrary labeled graphs. 

\begin{openproblem}
  Find dynamic $\DynFO$-programs for maintaining reachability in products of restricted classes of labeled graphs.
\end{openproblem}

Another interesting direction is to exhibit dynamic programs for other, more expressive query languages.

\begin{openproblem}
  Identify further expressive query languages that can be maintained dynamically.
\end{openproblem}

A candidate query language to be studied are nested regular expressions (NREs) \cite{PerezAG10}. NREs allow to express queries with some branching capabilities. For example, the NRE $(a[b])^*$ selects pairs of nodes that are connected by an $a^*$-labeled path such that every node on this path has an outgoing edge with label $b$. This query can easily be maintained in \DynFO, as it is bounded first-order reducible to reachability.  On the other hand, it is already unclear whether the query $(a[bc])^*$ can be maintained in \DynFO. 

 \bibliography{bibliography}

  \shortVersion{
 \appendix
 \clearpage
     \addcontentsline{toc}{0}{Appendix}
     \renewcommand{\thetheorem}{\thesection.\arabic{theorem}}
     \writeAppendix
   }
\end{document}